\documentclass{aa}
\usepackage{graphicx}
\usepackage{txfonts}
\usepackage{hyperref}
\usepackage[normalem]{ulem}
\hypersetup{colorlinks = true, citecolor = {blue}, urlcolor= {blue}}

\newcommand{\msun}{{\rm M}_{\odot}}

\def\gapprox{\;\rlap{\lower 3.0pt                       
        \hbox{$\sim$}}\raise 2.5pt\hbox{$>$}\;}
\def\lapprox{\;\rlap{\lower 3.1pt                       
        \hbox{$\sim$}}\raise 2.7pt\hbox{$<$}\;}

\usepackage{threeparttable}

\newcommand{\be}{ \begin{equation} }
\newcommand{\ee}{\end{equation}}

\newcommand{\ben}{\begin{enumerate}}
\newcommand{\een}{\end{enumerate}}

\renewenvironment{thebibliography}[1]{\begin{oldthebibliography}{#1}\setlength{\parskip}{0ex}\setlength{\itemsep}{0ex}}{\end{oldthebibliography}}

\usepackage{threeparttable}
\usepackage{diagbox}
\usepackage{siunitx}
\usepackage{ltablex}
\makeatletter
\renewcommand*\aa@pageof{, page \thepage{} of \pageref*{LastPage}}
\makeatother

\usepackage[dvipsnames]{xcolor}

\definecolor{darkgreen}{RGB}{31, 207, 31}

\usepackage{supertabular}

\usepackage{booktabs}

\usepackage{threeparttable}
\usepackage{multirow}
\usepackage{booktabs}
\newcommand{\mesa}{\textsc{mesa}}
\newcommand{\mesarsp}{\textsc{mesa-rsp}}

\begin{document}
    \title{The light curve model fitting of LMC Cepheids: MESA-RSP versus Stellingwerf's code predictions}

    \author{M. Deka
          \inst{1},
          M. Marconi
          \inst{1},
          R. Molinaro
          \inst{1},
          G. De Somma
          \inst{1,2,3},
          A. Bhardwaj
          \inst{4},
          E. Trentin
          \inst{1},
          S. Deb
          \inst{5},
          T. Sicignano
          \inst{6,7,1,3},
          I. Musella
          \inst{1},
          V. Ripepi
          \inst{1},
          E. Luongo
          \inst{1},
          \and
          Shashi M. Kanbur
          \inst{8}
          }

   \institute{INAF-Osservatorio astronomico di Capodimonte, Via Moiariello 16, I-80131 Napoli, Italy\\
              \email{mami.deka@inaf.it}
              \and
              INAF-Osservatorio Astronomico d’Abruzzo, Via Maggini sn, 64100 Teramo, Italy. 
              \and
              Istituto Nazionale di Fisica Nucleare (INFN)—Sez. di Napoli, Compl. Univ. di Monte S. Angelo, Edificio G, Via Cinthia, I-80126, Napoli, Italy.
              \and
              Inter-University Centre for Astronomy and Astrophysics (IUCAA), Post Bag 4, Ganeshkhind, Pune 411 007, India.
              \and
              Department of Physics, Cotton University, Panbazar, Guwahati 781001, Assam, India.
              \and
              European Southern Observatory, Karl-Schwarzschild-Strasse 2, 85748 Garching bei München, Germany.
              \and
              Scuola Superiore Meridionale, Largo San Marcellino10 I-80138 Napoli, Italy.
              \and
              Department of Physics, State University of New York, Oswego, NY 23126, USA
              }

   \date{}
	
\newcommand{\mar}[1]{\textcolor{blue}{#1}} 
\newcommand{\mami}[1]{\textcolor{magenta}{#1}}

\abstract
   {A major challenge in modeling classical Cepheids is the treatment of convection, particularly its complex interplay with pulsation. This inherently three-dimensional process is typically approximated in one-dimensional (1D) hydrocodes using dimensionless turbulent convection (TC) free parameters. Calibrating these parameters is essential for reproducing key observational features such as light curve amplitudes, secondary bumps, and the red edge of the instability strip.}
   {In this work, we calibrate TC parameters adopted in the publicly available Modules for Experiments in Stellar Astrophysics – Radial Stellar Pulsations (\mesarsp) code, through the comparison with both observational data of classical Cepheids and stellar parameter constraints from the Stellingwerf code -- one of the few codes capable of replicating a wide range of observed features of the classical pulsators.}
   {We compute multi-band ($V$, $I$, and $K_{s}$) \mesarsp\ light curves for 18 observed Large Magellanic Cloud (LMC) Cepheids, using stellar parameters that were determined based on Stellingwerf's code. By fine-tuning the mixing-length and eddy viscosity parameters, we calibrate the TC treatment in \texttt{MESA-RSP}. We then compare the resulting period-luminosity (PL), period–radius (PR), and period–mass–radius (PMR) relations with the prediction of Stellingwerf's model.}
   {We successfully reproduce multi-band ($V, I$ and $K_{s}$- band) light curves for 18 observed LMC Cepheids using stellar parameters determined with Stellingwerf's code and fine-tuned the mixing-length and eddy viscosity parameters in \mesarsp. Our models yield PL, PR, and PMR relations consistent with the results of Ragosta et al. (2019). Interestingly, our results are broadly in agreement with Ragosta et al. (2019), though we explicitly identify distinct mass–luminosity (ML) relations for fundamental-mode (FU) and first-overtone (FO) Cepheids for the first time. This suggests that the macroscopic phenomena affecting the ML relation depend on the stellar mass itself and/or on the effective temperature range. 
   Our investigation is focused on the calibration of the TC parameters, but we did not find a single set of convective parameter values that could reproduce all the light curves. Additionally, no statistically significant correlation is found between the stellar properties, such as the effective temperature or the stellar mass, and the convection parameters, although subtle trends with period and effective temperature are noted. As for the inferred Cepheid distances, our application of the model-fitting technique yields reddened distance moduli, which are in good agreement with the ones reported by Ragosta et al. (2019). This is not surprising, given that we adopted the same input stellar parameters, with only minor differences in the adopted model-atmospheres. 
   }
   {}

   \keywords{Convection - turbulence - stars: pulsations - stars: variables: Cepheids}
   \titlerunning{The light curve model fitting of LMC Cepheids}
   \authorrunning{M.\ Deka et al.}

   \maketitle
   
\section{Introduction}
\label{sec:intro}
Classical Cepheids (hereafter Cepheids) are a class of intermediate-mass evolved ($\sim 3-13~\msun$) pulsating stars crossing the instability strip during the core helium-burning phase \citep{bono00b,ande16,espi22,muse22}. They are particularly renowned for their well-defined period-luminosity (PL) relation, which establishes them as excellent distance indicators \citep{leav08,leav12,reis22}. Beyond their critical role in distance measurement, Cepheids also serve as ideal stellar laboratories for testing and validating stellar structure, evolution and pulsation \citep[e.g.,][]{cass11, moro12,deka25}. Their pulsations provide additional insights into their stellar properties through detailed comparisons between theoretical predictions and observational characteristics, which make them key tracers of stellar populations \citep{bono05,somm21}. However, the reliability of stellar models depends heavily on the accuracy of the adopted input micro- and macro-physics \citep{somm24}.

Cepheids occupy the classical instability strip (IS) of the Hertzsprung–Russell diagram (HRD), alongside other classical pulsators such as RR Lyrae stars, Type II and Anomalous Cepheids, $\delta$ Scuti and SX Phoenicis stars. These stars exhibit periodic variabilities caused by regular changes in radius and luminosity, driven by temperature-dependent fluctuations in gas opacity \citep{cox74}. Their pulsations are mainly sustained by the $\kappa$-mechanism, with the $\gamma$-mechanism amplifying the driving process \citep{cate15}. The blue edge of the instability strip is related to the depth of the ionizing regions of the most abundant elements in the stellar envelope. In contrast, the occurrence of a red edge is due to the quenching action performed by convection on pulsation towards the redder zone of the color-magnitude diagram. But the coupling between convection and pulsation not only affects the location of the red edge but also modifies the pulsation amplitudes and the capability of models to reproduce in detail the observed light and radial velocity variations \citep[see e.g.,][]{marc13,marc17,bhar17a,paxt19,somm20,somm22}.
Convection is a key mechanism, besides radiation, for transporting energy and momentum, as well as mixing elements within stars. It thus plays a significant role in shaping the stellar internal structure, evolution, and pulsation stability of stars. Given the extremely large scale of stars and their high viscosity, convection in stars typically occurs at extremely high Reynolds numbers \footnote{Reynolds number is a dimensionless quantity that predicts whether the fluid flow will be laminar or turbulent by comparing inertial forces to viscous forces.}. As a result, stellar convection is predominantly characterized by a fully developed turbulent state. The theory of turbulent convection has been developed over the past 100 years, but, due to computational limitations, it remains challenging to establish a comprehensive theory that is fully able to explain the nature of convection in stellar models. \citet{meak08} estimated that a completely resolved simulation of stellar turbulence would require at least $\sim 60$ years, based on computing trends available at the time and under the assumption that Moore's law would continue to hold. As a consequence, turbulent convection - an inherently three-dimensional (3D) phenomenon has been incorporated in one-dimensional (1D) codes based on simplified theoretical considerations. This involves several dimensionless free parameters, also called turbulent convection (TC) free parameters. 

To study the stellar interior structure and the different evolutionary phases, the stellar models still heavily rely on the steady-state 1D convective formulation -- the so-called mixing length theory \citep[MLT;][]{bohm58,heny65}. There have been some recent additions in this regard e.g., the turbulent convection model by \citet{kuhf86,kuhf87}. 
The MLT approach is only suitable for calculating the stellar structure and quasi-static evolution of a star, as the timescale of stellar evolution is much longer than the timescale of the stellar convective motion. We need a time-dependent non-local theory of convection to study stellar pulsations, as the timescale of pulsation and convection is of the same order of magnitude.

As mentioned above, in classical pulsators with relatively cool effective temperature,  convection plays an important role, although not in the entire envelope, but in the ionization zones -- the region of instability mechanisms, where steep temperature gradients are present. This region becomes convectively unstable, and energy is transported by convection more efficiently. Furthermore, during pulsation, the ionization fronts move back and forth, and the interaction between convection and pulsation becomes crucial in exciting and, most importantly, damping the pulsations. Additionally, since this region is unstable, the behaviour of convection becomes more complex than in other environments. The importance of including convection in pulsation models is highlighted by several studies \citep[e.g.,][]{gehm92b,feut99, marc09}. In particular, to reproduce the observed red edge of the IS, the amplitude and secondary features of the light curves, non-linear convective pulsation models are required. Several nonlinear convective pulsation codes have been developed in the last decades, which include a non-local time-dependent treatment of convection \citep[see e.g.][and references therein]{stel82a, gehm92a, Bono99a, feut99, koll02, marc05, smol08}. 

Early hydrodynamical studies with the Florida–Budapest code highlighted the importance of turbulent convection in shaping pulsation dynamics. \citet{koll02} showed that double-mode behavior in Cepheids arises naturally when turbulent convection is included in their hydrodynamic model, the Florida–Budapest code. However, \citet{smol08a} later argued that this behavior is actually a consequence of neglecting negative buoyancy in their pulsation code. The Cepheid phase lags were subsequently revisited by \citet{robe07}, also using the Florida–Budapest code. \citet{koll11} investigated period doubling in RR Lyrae models for the first time, providing the theoretical basis for its later detection in Blazhko RR Lyrae stars by Kepler. This framework was further supported by \citet{smol12} through their study of BL Herculis stars. Extending this line of work, \citet{smol12,smol14} explored variations of the ``canonical'' parameter sets in Type II Cepheid models and uncovered a range of dynamical phenomena, including period doubling, amplitude modulations, beat pulsations, and even chaotic variability behaviors. These works were motivated by \citet{koll11}, who first demonstrated period doubling in RR Lyrae models after fine-tuning the $\alpha$ parameters in the Florida–Budapest code.

Building on this foundation, recent efforts with  Modules for Experiments in Stellar Astrophysics – Radial Stellar Pulsations \citep[\mesarsp;][]{paxt10,paxt13,paxt15,paxt18,paxt19,jerm23} have aimed to calibrate turbulent convection (TC) parameters against observations. For RR Lyrae (both fundamental RRab and first-overtone RRc) stars, \citet{kova23,kova24} successfully reproduced radial velocity curves, though matching detailed light curve morphologies remained more challenging. This discrepancy suggests the need for either refined calibration or improved convection modelling. In case of Cepheids, the effects of various control parameters in MESA on Cepheid evolutionary tracks were investigated in great detail by \citet{ziol24}. Recent studies by \citep{kurb23,kova23,kova24,das25} suggest that the turbulent parameter prescriptions require rigorous calibration with observational data to simultaneously reproduce the observed light and radial velocity curve morphologies—particularly the amplitude and the secondary features (such as bumps) in the light curves. However, it is important to note that these parameters do not significantly affect the mean magnitudes/mean PL relation/topology of the IS edges \citep{somm22,deka24,das24}. Interestingly, \citet{deka24} found that some models pulsate beyond the estimated traditional linear red edge, which may arise due to an uncalibrated TC parameter set. Alternatively, it could indicate that such pulsators genuinely exist but have not been observed due to their relatively short lifespan. Before drawing any conclusions, further tests are needed.

In this work, we aim to calibrate the free convective parameters adopted in the \mesarsp\ code against observational data for Cepheid variables. To do so, we have selected several observed Cepheids that have already been precisely modelled using Stellingwerf’s code with their stellar parameters published in \citet[][hereafter R19]{rago19}. Given the large number of free parameters involved in the \mesarsp\ tool, we adopt this approach to reduce the number of unknowns in our TC parameter calibration process.

The remainder of the paper is organized as follows: In Section~\ref{sec:st vs mesa}, we provide a brief comparative overview of the Stellingwerf and \mesarsp\ codes. Section~\ref{sec:data} presents the data and methodology used in this work. The results, along with a discussion of the effects of the TC parameters on the light curves, are presented in Section~\ref{sec:results}. Finally, Section~\ref{sec:conclusion} outlines our findings and discusses future prospects.

\section{MESA-RSP versus Stellingwerf code}
\label{sec:st vs mesa}
Both \mesarsp\ and Stellingwerf's codes are nonlinear convective pulsation hydrocodes that couple convection with the equations of radiation hydrodynamics through an additional equation for turbulent energy.
However, the mathematical formulation of \mesarsp\, particularly in its treatment of convection, differs from that of Stellingwerf's code (see below for the details of the parameters). Therefore, a one-to-one calibration of the TC parameters is not possible. A practical and reliable approach in calibrating both codes is to match them against the same observed stellar properties using identical input stellar parameters. In addition, a detailed analysis of the temporal and spatial behavior of turbulent energy can offer further insights into the weaknesses/strengths of the underlying turbulent energy equations. Such a comparative study, focused exclusively on the temporal and spatial characteristics of the turbulent energy equations in both Stellingwerf’s code and \citet{kuhf86} has already been carried out by \citet{gehnwin92}. Although the models developed by \citet{kuhf86} and Stellingwerf remain among the most widely used for simulating the observations of the variable stars, the literature still lacks a thorough quantitative comparison of the dimensionless convective parameters involved in these two codes.

Recently, \mesarsp\ has been made publicly available, which enables us to model linear/non-linear radial pulsations exhibited by variable stars like Classical and Type II Cepheids and RR Lyraes \citep{smol08,paxt19}. It implements the time-dependent turbulent convection model of \citet{kuhf86}, following the stellar pulsation framework described in \citet{smol08}. This convection model incorporates eight free dimensionless parameters:

\begin{enumerate}
    \item[$\alpha$] (Mixing length): Distance a convective element travels before dissipating.
    \item[$\alpha_m$] (Eddy viscosity): Effective viscosity from turbulent momentum transport.
    \item[$\alpha_s$] (Turbulence source): Turbulence generation via shear, buoyancy, or forcing.
    \item[$\alpha_c$] (Convective transport): Efficiency of energy transport by convection.
    \item[$\alpha_d$] (Turbulence dissipation): Conversion of turbulent energy into heat.
    \item[$\alpha_p$] (Turbulent pressure): Pressure contribution from turbulent motion.
    \item[$\alpha_t$] (Turbulent flux): Transport of mass/momentum/energy by turbulence.
    \item[$\gamma_r$] (Radiative losses): Energy lost from convective eddies via radiation.
\end{enumerate}
Four different combinations of these parameters are given in \citet{paxt19}, labeled as A, B, C, and D. In set A, turbulent 
pressure, turbulent flux, and radiative cooling are neglected (i.e., $\alpha_{p}=\alpha_{t}=\gamma_{r}=0$) and values of 
$\alpha_{s},\alpha_{c}$ and $\alpha_{d}$ are kept the same as in Table~\ref{table:convective_set}. Under this condition, the 
time-independent version of the \citet{kuhf86} model becomes locally equivalent to the standard mixing-length theory 
\citep[MLT;][]{bohm58}. Set B adds radiative cooling. This parameter was not present in the original \citet{kuhf86} model and 
is set to $2\sqrt{3}$ following \citet{wuch98}. Set C adds turbulent pressure ($\alpha_{p}$) and convective flux 
($\alpha_{c}\equiv\alpha_s$) \citep{yeck98}. The $\alpha_p$ parameter was implicitly included in the \citet{kuhf86} model through turbulent convection, and thus turbulent pressure, but it had a fixed value of $\frac{2}{3}$. \citet{yeck98} extended the model by allowing $\alpha_p$ to be scaled. The $\alpha_c$ parameter, which was not present in the original \citet{kuhf86} model, was also introduced by \citet{yeck98}.  Set D includes all these effects simultaneously \citep{paxt19}. Parameter values follow Tables 3--4 of \citet{paxt19}, summarized in Table~\ref{table:convective_set} in the Appendix~\ref{app:fiducialparameter}. For a more detailed discussion on these parameters, we refer the reader to \citet{kuhf86,wuch98,yeck98,smol08}.

The original version of Stellingwerf's code was developed by \citet{stel82a} based on the unpublished work of Castor (1968)  and was later improved by \citet{bono94,bono99,somm24}. The model includes a free convective parameter (analogous to the $\alpha$'s in \mesarsp) proportional to the ratio between the mixing length \(\lambda\) that is the path covered by the convective elements and the pressure height scale. This parameter in Stellingwerf's code is used to close the nonlinear equation system. Another free parameter is the eddy-viscosity dissipation parameter or eddy-viscosity mean free path. Both the turbulent pressure and
the eddy viscosity pressure, ignoring a constant of order unity,
form one component of a nine-component symmetric tensor
(Reynolds stress tensor). The third free parameter is 
 the turbulent diffusion parameter \(\lambda_{ovs}\). Although the relations between these parameters have not been studied in detail through comparisons between theoretical observables and observational constraints, $\lambda_{ev}$ and $\lambda_{ovs}$ are typically chosen such that $\lambda_{ev} \approx \lambda$ and $\lambda_{ovs} \approx \lambda/10$, reducing the number of free parameters that need to be calibrated. For a detailed description, we refer the readers to \citet{bono94}.

\section{Data and Methodology}
\label{sec:data}
\subsection{The sample and the selected input model parameters}
\begin{figure}
\includegraphics[width=0.45\textwidth,keepaspectratio]{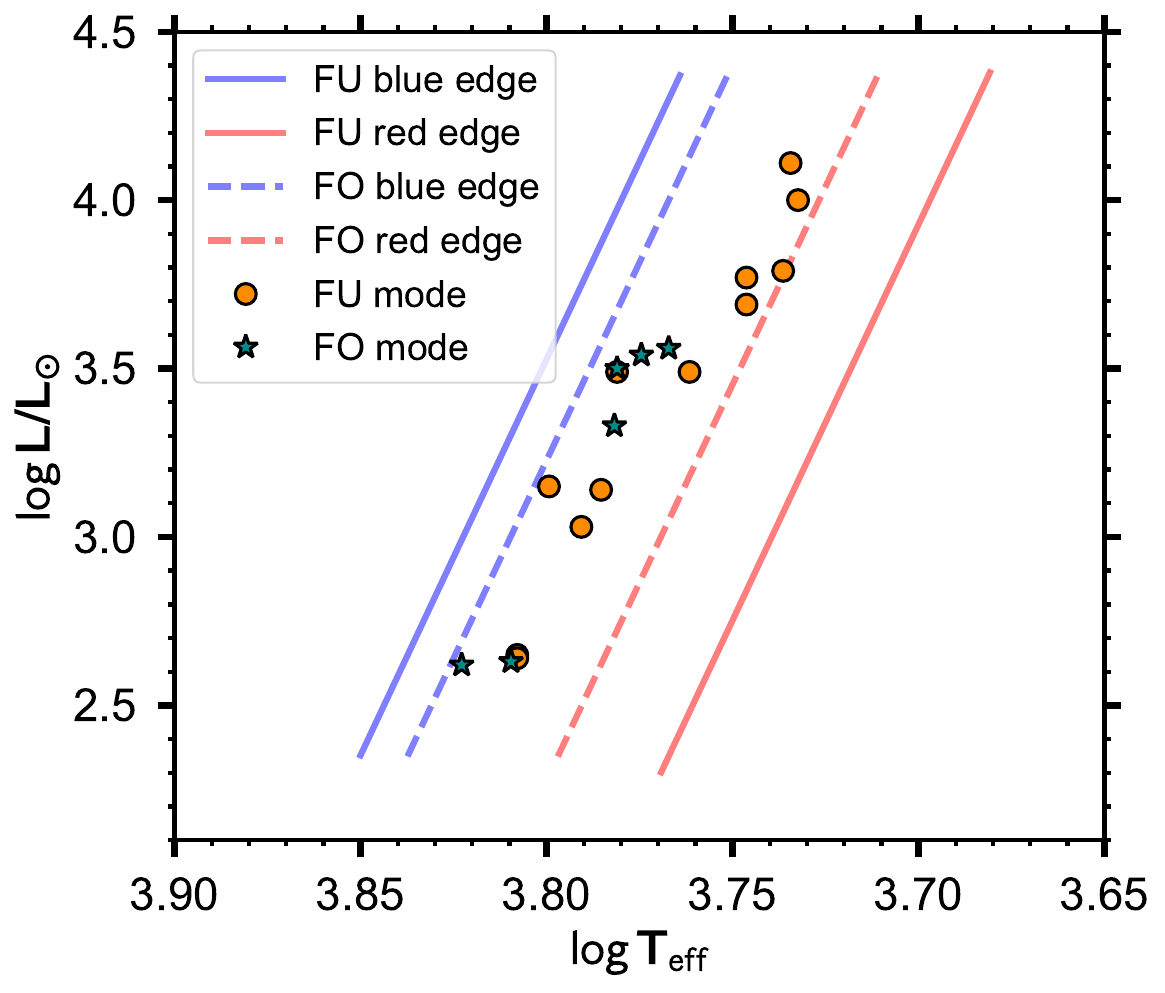}
\caption{Location of the Cepheids selected in this work on the HRD. The IS edges are taken from \citet{deka24}. }
\label{fig:hrd}
\end{figure}

We have selected 18 Large Magellanic Cloud (LMC) Cepheids from R19, consisting of 11 fundamental (FU) and 7 first-overtone (FO) mode Cepheids. The periods of these Cepheids span a wide range, approximately \(1\) to \(30\) days. This sample also spans a wide range of luminosities and shapes of light curves. The stellar parameters required for pulsation model computations—mass (\(M\)), luminosity (\(L\)), effective temperature (\(T_{\rm eff}\)), and chemical composition (\(X, Y,\) and \(Z\))—are also taken from R19. Additionally, we adopted the mixing-length efficiency parameter (\(\alpha\)), one of the TC parameters, from this paper as our starting point. The list of parameters is reported in Table~\ref{tab:input_cepheids}, whereas the position of these Cepheids in the HRD is shown in Fig.~\ref{fig:hrd}.

We have used multi-filter light curves in the $V,I,$ and the Visible and Infrared Survey Telescope for Astronomy (VISTA) $K_{s}$-bands of the selected Cepheids to calibrate the TC parameters.  
The observed optical $V$- and $I$-band light curves were obtained from the Optical Gravitational Lensing Experiment (OGLE)-IV database\footnote{\url{https://ogle.astrouw.edu.pl/}} \citep{sosz15}. The near-infrared $K_{s}$-band light curves were sourced from the ``VISTA near-infrared $Y, J, K_s$ survey of the Magellanic Clouds system" (VMC) survey \citep{cion11, ripe16, ripe17, Ripepi2022}.

In \mesarsp, there are four different combinations of the eight free TC parameters, which are considered fiducial and have not been calibrated to observations. For the initial computation, we set the mixing-length efficiency parameter as given in R19, while the remaining seven parameters (\(\alpha_m, \alpha_c, \alpha_s, \alpha_d, \alpha_p, \alpha_t\), and \(\gamma_r\)) were taken as listed in Table~\ref{table:convective_set}.  

\mesarsp\ solves the pulsation equations using a Lagrangian mesh, structured into inner and outer zones, based on a specified anchor temperature, which we set to \(T_{\rm{anchor}} = 11,000\) K. Initially, the total number of Lagrangian mass cells was set to \(N_{\rm total} = 200\), with \(N_{\rm outer} = 60\) cells above the anchor. However, if this configuration failed to generate the initial model, we adjusted the setup to \(N_{\rm total} = 150\) and \(N_{\rm outer} = 40\). The selection of numerical parameters, including the inner boundary temperature (\(T_{\rm{inner}} = 2 \times 10^6\) K), follows the guidelines as outlined in \citet{paxt19} (for further details, see Appendix~B of \citet{deka24}). These parameters remain fixed throughout the entire calibration process.  

To perturb the initial model, we set the surface velocity parameter, RSP\_kick\_vsurf\_km\_per\_sec, to 0.1. For mode selection, RSP\_fraction\_1st\_overtone was set to 0 for FU mode and 1 for FO mode. The RSP\_fraction\_2nd\_overtone was set to 0 in all cases, as our sample consists exclusively of FU and FO Cepheids. If we were unable to obtain the desired pulsation mode for a particular star, we systematically adjusted the RSP\_kick\_vsurf\_km\_per\_sec parameter (discussed in more detail in Sec.~\ref{sec:3.2}) \footnote{The \mesarsp\ inlist used to compute the models in this work is available at \url{https://github.com/mami-deka/LNA_CMD}}. Such difficulties typically arise in the hysteresis domain, where the resulting mode depends sensitively on the initial conditions \citep{koll02,smol10}.

We consider the model to reach full amplitude stable mode when the pulsation period $P$, the fractional growth of the kinetic energy per pulsation period, and the 
amplitude of radius variation $\Delta R$ do not differ by more than $\sim 0.0001$ over the last $\sim 100$-cycles of the total integrations computed. It takes 
$\sim 1000-4000$ pulsation cycles to reach full-amplitude stable mode pulsations for these models.

\begin{table}[htbp]
\caption{List of Cepheids analysed in this work.}
\label{tab:input_cepheids}
\begin{tabular}{lccccccc}
\toprule
ID & P & Mode & M ($M_{\odot}$) & $\log(L/L_{\odot})$ & $T_{\text{eff}}$ & $\alpha_{\text{ML}}$ \\
 & (days) &  & &  (dex) &  (K) & \\
\midrule
1481 & 0.922 & FO & 3.00 & 2.62 & 6650 & 1.50 \\
3131 & 1.095 & FO & 2.80 & 2.63 & 6450 & 1.80 \\
3004 & 1.524 &FU& 3.00 & 2.65 & 6425 & 1.60 \\
1523 & 1.572 &FU& 2.80 & 2.64 & 6425 & 1.60 \\
3113 & 2.068 & FO & 5.20 & 3.15 & 6300 & 1.60 \\
2138 & 3.011 &FU& 3.80 & 3.03 & 6175 & 1.70 \\
3105 & 3.514 & FO & 4.80 & 3.33 & 6050 & 1.60 \\
0961 & 3.711 &FU& 3.50 & 3.14 & 6100 & 1.50 \\
1475 & 4.387 & FO & 5.60 & 3.50 & 6040 & 1.70 \\
1124 & 4.457 &FU& 5.60 & 3.49 & 6040 & 1.70 \\
1310 & 5.126 & FO & 5.70 & 3.54 & 5950 & 1.49 \\
0813 & 5.914 & FO & 5.70 & 3.56 & 5850 & 1.53 \\
2012 & 7.458  &FU& 5.30 & 3.49 & 5775 & 1.60 \\
1954 & 12.950 &FU& 5.30 & 3.69 & 5575 & 1.70 \\
0546 & 15.215 &FU& 5.20 & 3.77 & 5575 & 1.70 \\
1086 & 17.201  &FU& 5.70 & 3.79 & 5450 & 1.50 \\
2944 & 20.320 &FU& 6.50 & 4.00 & 5400 & 1.70 \\
2019 & 28.103 &FU& 7.70 & 4.11 & 5425 & 1.70 \\
\bottomrule
\end{tabular}%
\tablefoot{The table reports the Cepheids used in this study, along with the best-fitting model mass, luminosity, effective temperature, and mixing length parameter obtained by R19 using the Stellingwerf code. The chemical compositions are set to $X=0.742,Y=0.25$ and $Z=0.008$ except for 2019 ($X=0.692,Y=0.30$ and $Z=0.008$) similar to R19.}
\end{table}

\subsection{Conversion of bolometric light curves to $V,\  I$ and $K_{s}$-filter}
Pulsation codes typically output bolometric light curves that need to be converted into observational photometric systems to allow for a direct comparison with the observed curves. While the most rigorous approach involves solving the radiative transfer equation --  accounting for local thermodynamic equilibrium (LTE) or non-LTE (NLTE) effects-- to compute filter-specific magnitudes, this is computationally expensive. For practical applications, pre-computed bolometric correction (BC) tables offer an efficient alternative.
In this work, we use \mesa’s built-in interpolation routines to extract BCs as a function of stellar parameters (e.g., $T_{\rm eff}, \log{g}$, [Fe/H]) at each pulsation phase. We applied BCs to all phases of the light curve to account for depth differences across different pulsation cycles.

We have adopted the BC tables from \citet{leje98} for optical bands $V$ and $I$. For the $K_{s}$-band, we have used the BC tables from the MESA Isochrones \& Stellar Tracks (MIST) model grids \citep{choi16,dott16}\footnote{\url{https://waps.cfa.harvard.edu/MIST/model_grids.html}}, which are derived from the C3K grid (Conroy et al., in prep.). The C3K grid employs 1D atmospheric models based on ATLAS12/SYNTHE \citep{kuru70,kuru93}.

\subsection{TC parameter calibration}
\label{sec:3.2}
We computed models for each target using four distinct combinations of fiducial TC parameters (sets A, B, C, and D). For each target, we selected the best-fitting model and then refined it by varying either $\alpha$ or $\alpha_m$ in small steps of 0.1 and 0.01, respectively, either increasing or decreasing them as needed to match the observed amplitudes. It should be noted that $\alpha$ and $\alpha_m$ are not entirely independent: changing $\alpha$ also affects $\alpha_m$ through the underlying convection equations. It has been discussed in details in \citet{koll02}. This interdependence is typical of most 1D pulsation codes.

\begin{figure}
\includegraphics[width=0.45\textwidth,keepaspectratio]{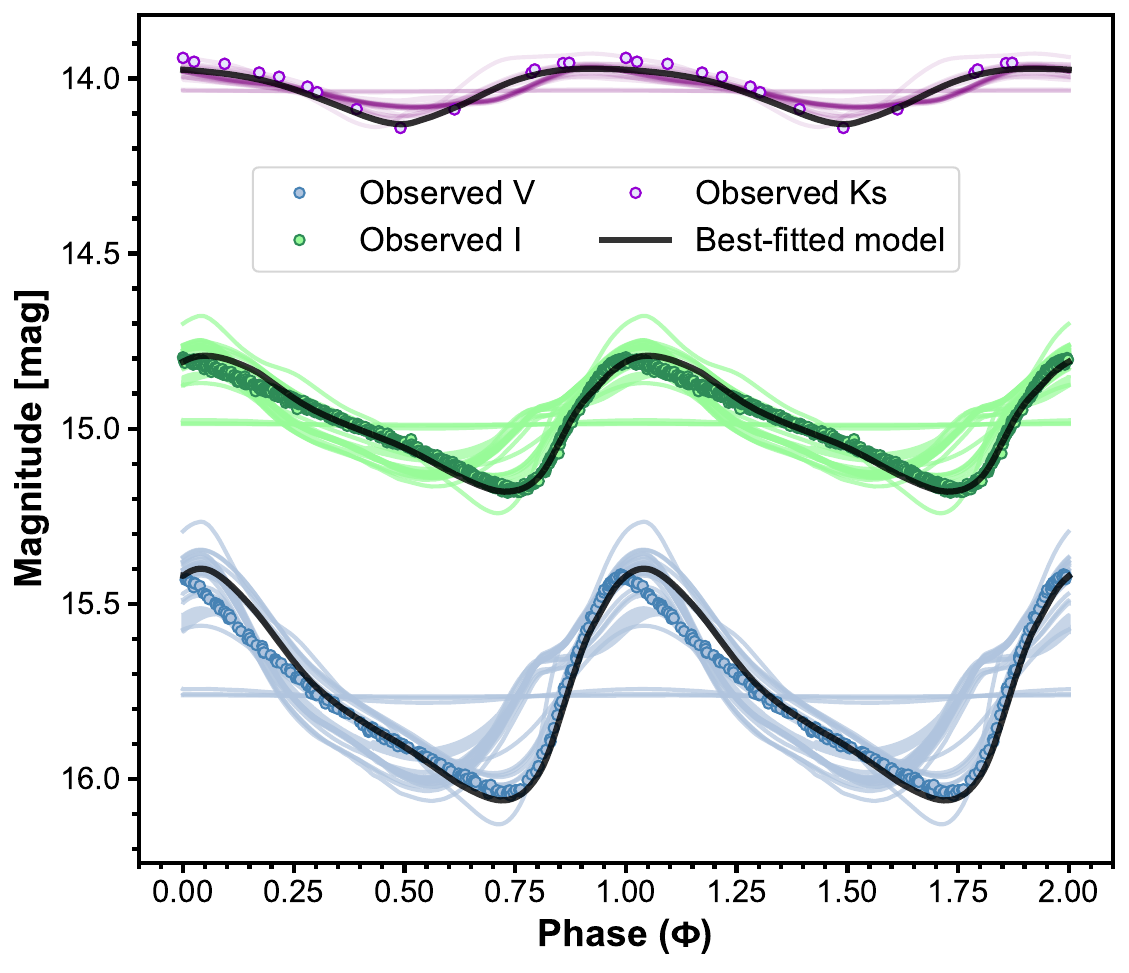}
\caption{The observed light curve (data points) for OGLE-LMC-CEP-1124 star is shown along with the best-fitting model predictions overlaid on all the computed models for this star. The $\chi^2$ values of the models range from 1.3 (best-fit model) to 49.4 (worst-fit model).}
\label{fig:all_model}
\end{figure}

The surface velocity parameter, RSP\_kick\_vsurf\_km\_per\_sec, was set to 0.1 (same as the \mesarsp\ fiducial parameter) in most cases. However, in instances where the initial setting did not produce the desired pulsation mode, we adjusted this parameter in steps of 0.5. Additionally, in a few cases where the model period differed significantly from the observed period, we made slight modifications to \(L\) and \(T_{\rm eff}\).  
For each target, we computed approximately 50 models (up to 100 in some cases) and identified the best-matched model using the model-fitting technique described in the next section. An example of all possible models for a Cepheid with the best fitted one is shown in Fig.~\ref{fig:all_model}.

\subsection{Best-fit model estimation}
The best-fit model was determined using a model-fitting technique following \citet{marc17,rago19}, implemented as follows:

\subsubsection{Light Curve Phasing}
First, the observed light curves were phased using:
\begin{equation}
\label{eq:phase}
\Phi = \frac{t - t_{0}}{P} - \text{Int}\left(\frac{t - t_{0}}{P}\right),
\end{equation}
where $t_{0}$ is the epoch of maximum light in the $V$-band, $P$ is the pulsation period (in days), and $t$ denotes observation times. Both theoretical and observed light curves ($V$, $I$, and $K_{s}$-bands) were then phase-aligned such that maximum brightness corresponds to phase zero.

\subsubsection{Interpolation and $\chi^2$ Minimization}
Theoretical light curves were interpolated to match the observed phases using a cubic spline (implemented via Python's \texttt{scipy.interpolate}). The $\chi^2$ function, defined as:
\begin{equation}
\label{eq:chi_fun}
\chi^2 = \frac{1}{N_{\rm bands}}\sum_{i=1}^{N_{\text{bands}}} \frac{1}{N^{j}_{DOF}} \sum_{j=1}^{N_{\text{points}}}\frac{\left[ m_j^i - \left( M_{\text{model}}^i (\phi_j^i + \delta\phi^i) + \delta M^i \right) \right]^2}{\sigma^j_i},
\end{equation}
was minimized using the L-BFGS-B optimizer\footnote{L-BFGS-B is a limited-memory technique to solve large-scale nonlinear optimization problems with simple variable bounds.} \citep{byrd95}. Here:
\begin{itemize}
\item $i$: band index ($V$, $I$, $K_{s}$);
\item $j$: data point index;
\item $N_{\rm bands}$: number of bands;
\item $N^{j}_{DOF}$: $N_{\rm points}-2$; number of degrees of freedom;
\item $\delta M^i$: magnitude shift/distance modulus for $i$th band;
\item $\delta\phi^i$: phase shift between model and observed light curves;
\item $M_{\text{model}}$, $m$: model and observed apparent magnitudes, respectively;
\item $\sigma^i_j$: observational uncertainty.
\end{itemize}

\subsubsection{Bayesian Refinement}
The L-BFGS-derived parameters ($\delta M^i$, $\delta\phi^i$) and their uncertainties were further refined through Markov Chain Monte Carlo (MCMC) analysis using the \textsc{emcee} - Python package \citep{fore13}. We have used 32 walkers, 10000 iterations and 2000 burn-ins for the MCMC sampling.

\section{Results and Discussion}
\label{sec:results}
\subsection{Light curve with MESA-RSP fiducial TC parameters}
We initially computed a set of models for each target by setting one of the eight TC parameters in \mesarsp, \(\alpha\), to the same value as given by R19, while keeping the remaining seven parameters at their fiducial values as defined in \citet{paxt19}. We have found that sets A and B produced nearly identical results in terms of morphological features. However, the models from set A yielded periods and amplitudes much closer to the observed values than those from set B. Similarly, sets C and D provided comparable results. In some cases, observations were better matched with models from set C or set D rather than from sets A or B.  
The \(T_{\rm eff}\) of the stars played a crucial role in determining the best-matching set. Specifically, stars with \(T_{\rm eff} > 6000\) K were better matched with set A models, whereas those with \(T_{\rm eff} < 6000\) K showed better agreement with either set C or set D. This is expected as the convection efficiency increases towards the red edge (lower temperature). Furthermore, for some stars, we were initially unable to obtain the desired pulsation modes using the fiducial parameter settings.  However, after adjusting the surface velocity parameter and keeping other parameters fixed, we successfully obtained the desired pulsation mode for these stars. Overall, 11 stars are best matched with set A, 1 with set C, and 6 with set D using the fiducial parameters based on the morphological structure but not amplitudes.

\subsection{Best-fitted models}
\begin{figure*}
\includegraphics[width=0.95\textwidth,keepaspectratio]{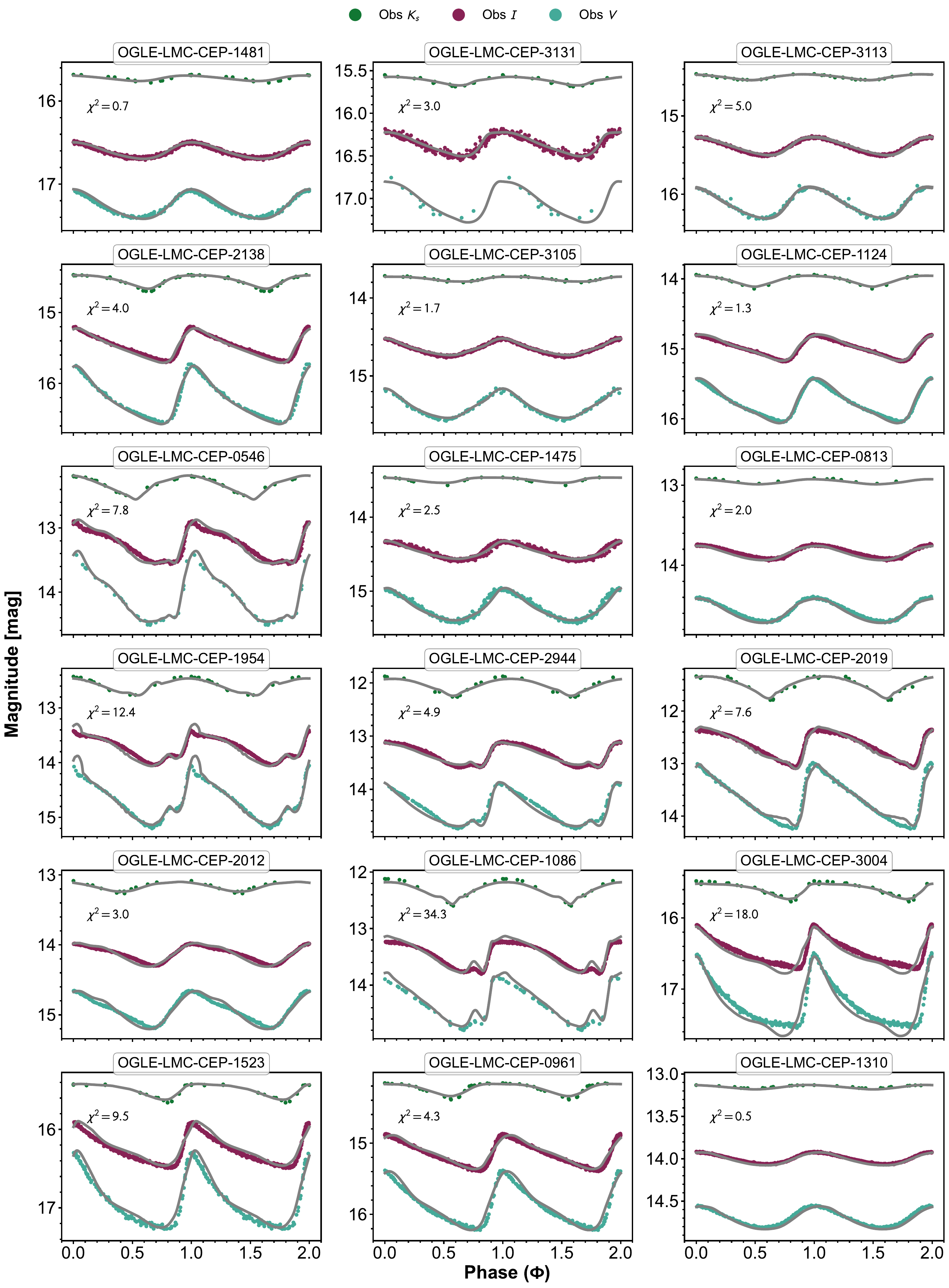}
\caption{Observed light curves (data points) are shown for the selected Cepheids with the best-fitting model predictions overplotted as continuous lines. }
\label{fig:bestfit1}
\end{figure*}

\begin{figure*}
\includegraphics[width=0.95\textwidth,keepaspectratio]{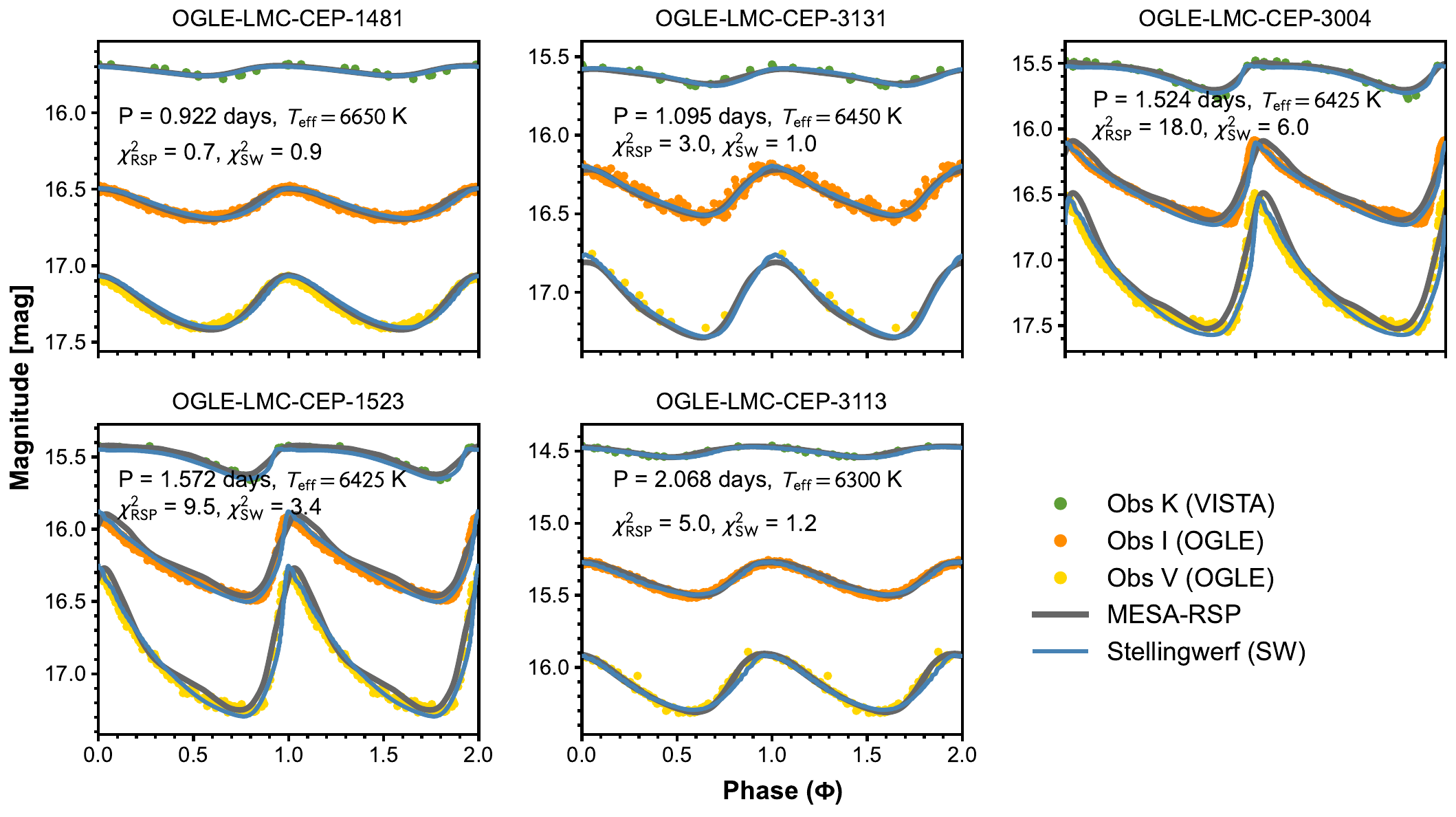}
\caption{Observed light curves (data points) are shown for five selected Cepheids with the best-fitting model predictions from \mesarsp\ and Stellinwerf code over-plotted as continuous lines. }
\label{fig:bestfit1_sw}
\end{figure*}

\begin{figure*}
\includegraphics[width=0.9\textwidth,keepaspectratio]{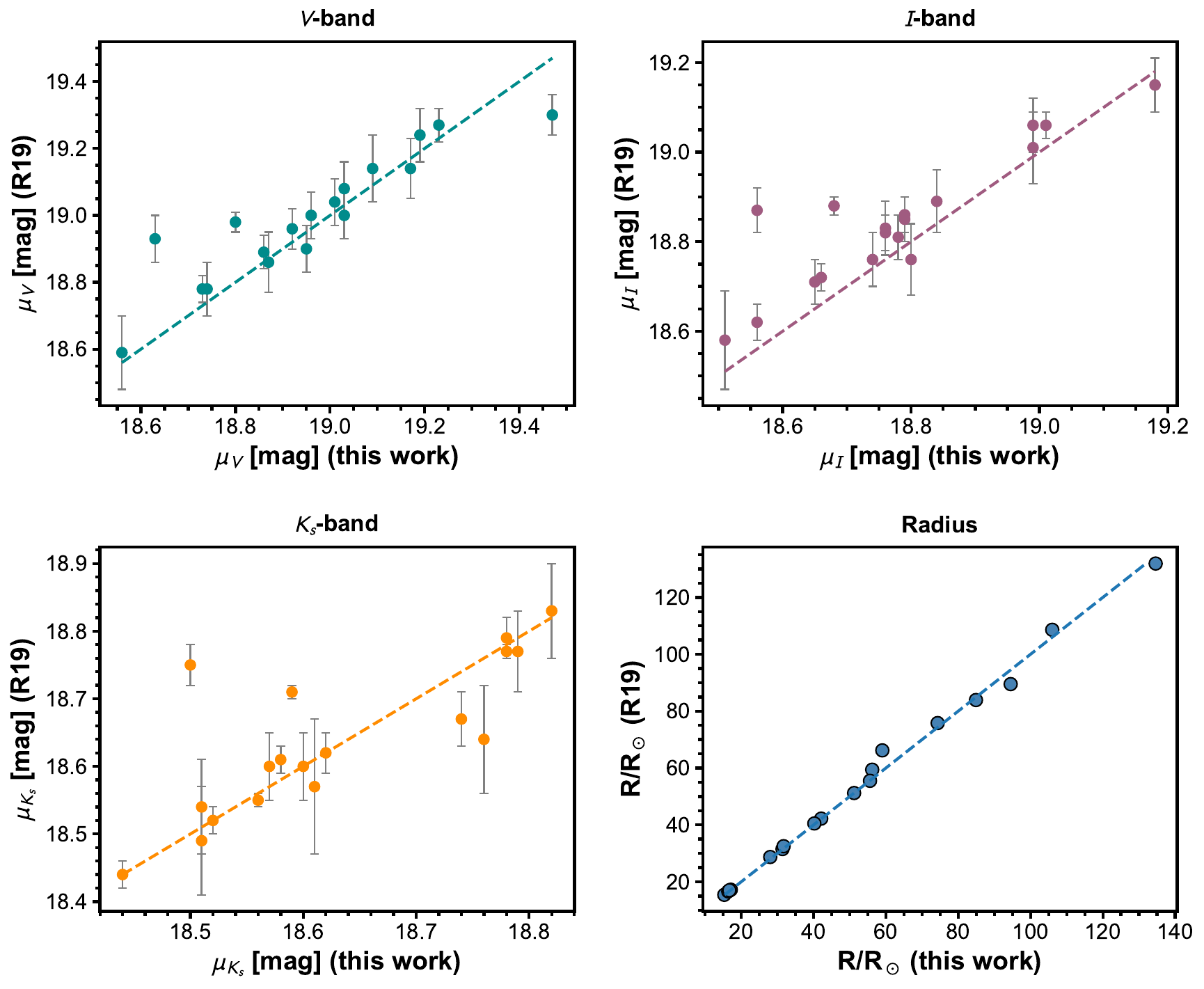}
\caption{This plot presents a comparison between the reddened distance moduli (in the $V, I$, and $K_{s}$ bands) and stellar radii derived in this work with those reported by R19. Each panel shows the respective comparison, including error bars on the distance moduli, showing the consistency and deviations across photometric bands and radius estimates.
}
\label{fig:mag_shift}
\end{figure*}

The morphology of the \mesarsp\ model with the fiducial TC parameters shows good agreement with the observed light curves both in terms of shape and the positioning of small bumps. However, the model amplitudes vary from approximately half to twice the measured amplitude and need to be matched. To achieve this, we varied the parameters $\alpha$ and $\alpha_m$ individually to adjust the amplitudes in accordance with the observations. Additionally, we obtained the best-fit parameters ($\Delta \phi$ and $\Delta m$) along with their associated uncertainties using the model-fitting technique described in Sec.~\ref{sec:data}. The observed light curves along with the best-matched models and for a subset of Cepheids with Stellingwerf's best-fit curves, are shown in Fig.\ref{fig:bestfit1} and \ref{fig:bestfit1_sw}, respectively. 
We find a very good global agreement between the theoretical curves produced with \mesarsp\ and Stellingwerf's code. Fig.~\ref{fig:corner} shows a representative posterior distribution from the Bayesian refinement analysis for the star OGLE-LMC-CEP-1124, demonstrating the parameter covariances and final uncertainty estimates. The best-fit parameters (stellar parameters and TC parameters) are summarized in Table~\ref{tab:bestfit_params}. In Fig.~\ref{fig:mag_shift}, we present the visual (top left panel), the I band (top right panel) and the unreddened (bottom left panel) distance modulii together with the radii (bottom right panel) derived in this work for each star, alongside the corresponding values from R19 for comparison. The radii show good agreement with those reported by R19. However, we observe subtle differences in the distance moduli for some stars. This small difference is likely due to differences in the bolometric corrections and color-temperature transformations or the methods employed to perform these conversions. Despite the consistency in the underlying models, this highlights that the choice of model atmospheres and their interpolation can have a significant impact on the derived distances. The phase shifts are found to range from approximately $-0.02$ and $0.08$. 

We obtained the highest number of observation-matched models from set A, where $\alpha_p = \alpha_t = \gamma_r = 0$. To further investigate the impact of including these parameters on those models, we increased $\alpha_p$ in steps of 0.1 from 0.0 to 1.0, set $\gamma_r$ to either 0 or 1 and $\alpha_t = 0.01$. We refer to this new set of models as the ``set D equivalent". Examples of these models, along with those from set A, are shown in Fig.~\ref{fig:setA-setD_comp}. These models also provided a good match with the observations; however, in order to reproduce the observed amplitudes, it was necessary to again tune either $\alpha$ or $\alpha_m$ compared to the best-fit models from set A. This is because $\alpha_p$ adjusts the turbulent pressure, which directly influences the work integral. At the same time, the work integral is affected by the eddy viscosity, which is scaled by $\alpha_m$. Therefore, both parameters must be adjusted together to keep the work integral unchanged. In general, scaling one alpha parameter will require a corresponding adjustment of the others. Furthermore, in most of the stars, $\alpha_p$ could not be increased beyond 0.2. Higher values either resulted in negative growth rates or failed to reproduce the observed light curve morphology. 

We also found it necessary to modify the surface velocity parameter, since the values used for set A did not yield the desired pulsation modes. We achieved a good match for nearly all the stars. In some models, the inclusion of these parameters improved the light curve morphology, resulting in a better agreement with observations. One such example is OGLE-LMC-CEP-3131 (see top-center-left panel of Fig.\ref{fig:setA-setD_comp}). However, in some cases, the fit worsened. For instance, in OGLE-LMC-CEP-2944, although the overall match remained reasonable, the characteristic bump disappeared (see middle-right panel of Fig.\ref{fig:setA-setD_comp}). Setting $\gamma_r$ to 1 could reproduce a light curve identical to that of set A, but it changes both the amplitude and the morphology of the radial velocity curves. The best-fit parameters (stellar and TC parameters) for this set of TC parameters are summarized in Table~\ref{tab:bestfit_params_setD}.

As discussed earlier, similar light curve shapes can be reproduced with different combinations of TC parameters, leading to some degeneracy in the model outputs. One way to help reduce this degeneracy is to include radial velocity data in the fitting process. While different parameter combinations such as mixing length and eddy viscosity, can produce comparable light curve amplitudes, they often lead to distinct radial velocity amplitudes. This occurs because the light curve amplitude primarily reflects temperature variations near the photosphere, whereas the radial velocity is governed by the dynamical response of the stellar envelope. For example, increasing the eddy viscosity can suppress the mechanical motion of the pulsation without significantly affecting the temperature contrast, allowing two models to produce similar light curves but differing radial velocity behavior. We note, however, that this effect is most relevant for $V$ and bluer bands, which are sensitive to $T_{\rm eff}$ variations, while infrared bands, such as $K_s$, are more sensitive to radius changes and therefore not entirely independent of the radial velocity data. Furthermore, as the 1D convection models rely on strong simplifications and artificial parameterizations, a parameter set that reproduces the observed light curve most closely may not necessarily be more physically accurate than another.

To illustrate this, we present a comparative analysis between the set D equiv and the set A models. While the period, light curve morphology, and amplitude closely match for most stars in our sample, we observe notable differences in the surface velocity amplitudes, as shown in Fig.~\ref{fig:setA-setD_ample_comp}.
However, we acknowledge that a more systematic investigation is required to quantify these effects more robustly. We plan to address this in a future study.

\subsection{Correlation between TC parameters and stellar properties}
\begin{figure*}
\includegraphics[width=0.9\textwidth,keepaspectratio]{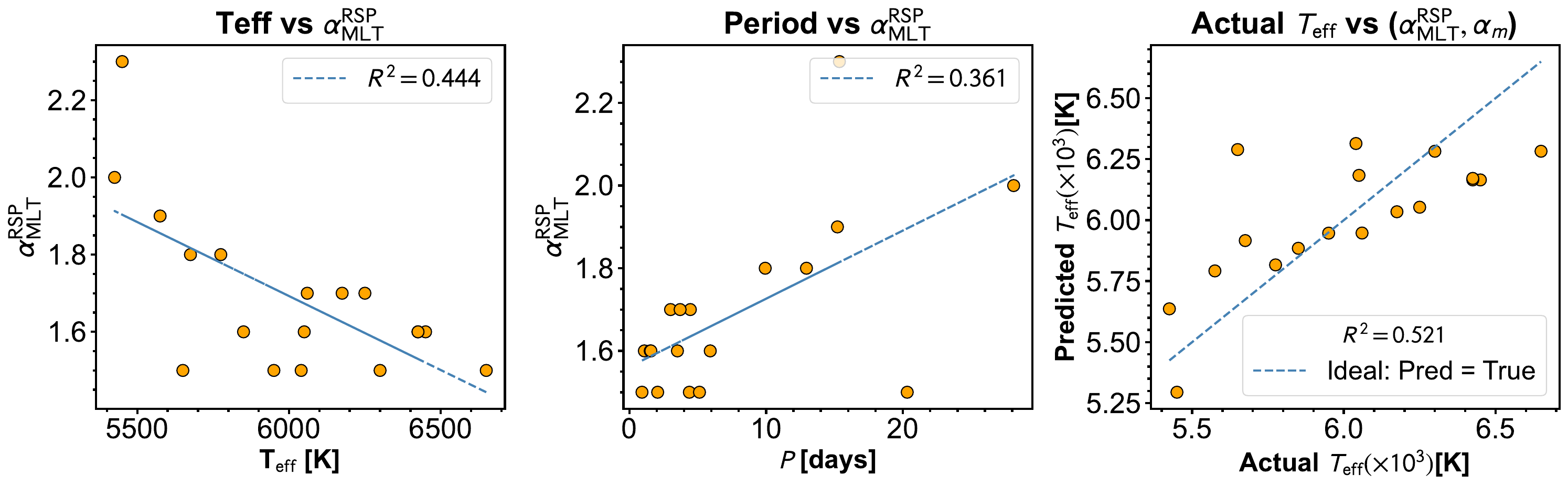}
\caption{Correlations between the stellar parameters and the TC parameters (\( \alpha^{\rm RSP}_{\rm MLT} \) and \( \alpha_m \)) are shown for cases with \( R^2 > 0.3 \). The left and middle panels display linear regressions (red dashed lines) along with the corresponding \( R^2 \) values. Among all comparisons, only the relationships between \( \alpha^{\rm RSP}_{\rm MLT} \) and effective temperature or period, as well as between \( \alpha^{\rm RSP}_{\rm MLT} \), \( \alpha_m \), and \( T_{\rm eff} \), exhibit moderate correlations. Most other parameter combinations show weak or no significant correlation.
}
\label{fig:mlt_vs_tc}
\end{figure*}

To systematically investigate potential relationships between the TC parameters and global stellar properties, we performed a series of linear regressions between key quantities. We focused on two key TC parameters: $\alpha^{\rm RSP}_{\rm MLT}$  and $\alpha_m$, comparing them with \( T_{\rm eff} \), mass, luminosity, and pulsation period. A few examples of these comparisons is presented in Figure~\ref{fig:mlt_vs_tc}.

We find that the majority of parameter combinations show no strong or statistically robust correlations, with $R^2$ values typically below 0.3. This suggests that neither $\alpha^{\rm RSP}_{\rm MLT}$ nor $\alpha_m$ exhibits a strong linear dependence on the basic stellar parameters within our sample. The most notable exception is a moderate positive correlation between \( \alpha^{\rm RSP}_{\rm ML} \) and both \( T_{\rm eff} \) and the pulsation period. The trend in temperature suggests that cooler stars tend to require higher convective efficiency, as expected and reported in previous literature \citep{somm22}. Similarly, the trend with period may imply an indirect connection with stellar radius or evolutionary stage, as longer-period variables typically correspond to more evolved stars.

In contrast, correlations between $\alpha_m$ and parameters such as $T_{\mathrm{eff}}$, mass, luminosity, and period are weak or statistically insignificant.
However, for RR Lyrae stars, \citet{kova23,kova24} reported a significant relationship between $\alpha_m$ and $T_{\mathrm{eff}}$. The discrepancy between their findings and ours may stem from the degeneracy between $\alpha_m$ and $\alpha^{\mathrm{RSP}}_{\mathrm{ML}}$. In their work, $\alpha^{\mathrm{RSP}}_{\mathrm{ML}}$ was held fixed while $\alpha_m$ was varied, whereas in our analysis, $\alpha^{\mathrm{RSP}}_{\mathrm{ML}}$ was set according to R19 -- where it varies across stars -- before adjusting $\alpha_m$ to match the observed amplitudes. This methodological difference likely contributes to the weaker trends observed in our results.
To explore this further, we examined the correlation between $\alpha_m$ and $\alpha^{\mathrm{RSP}}_{\mathrm{ML}}$, finding only a weak, scattered relationship ($R^2 = 0.029$). While this suggests a possible empirical connection, the large intrinsic scatter points to substantial underlying variability. Additionally, the absence of stronger trends may partly reflect the limited sample size, which reduces the statistical power of our analysis. Another possibility is that such a correlation may not exist for Cepheids at all. This weaker parameter dependence could be due to their convective layers being more extended than in RR Lyrae stars. Future studies with larger samples and broader parameter coverage will be essential to clarify these relationships.

\subsection{Stellar global parameter correlations }
\begin{figure*}
\includegraphics[width=0.9\textwidth,keepaspectratio]{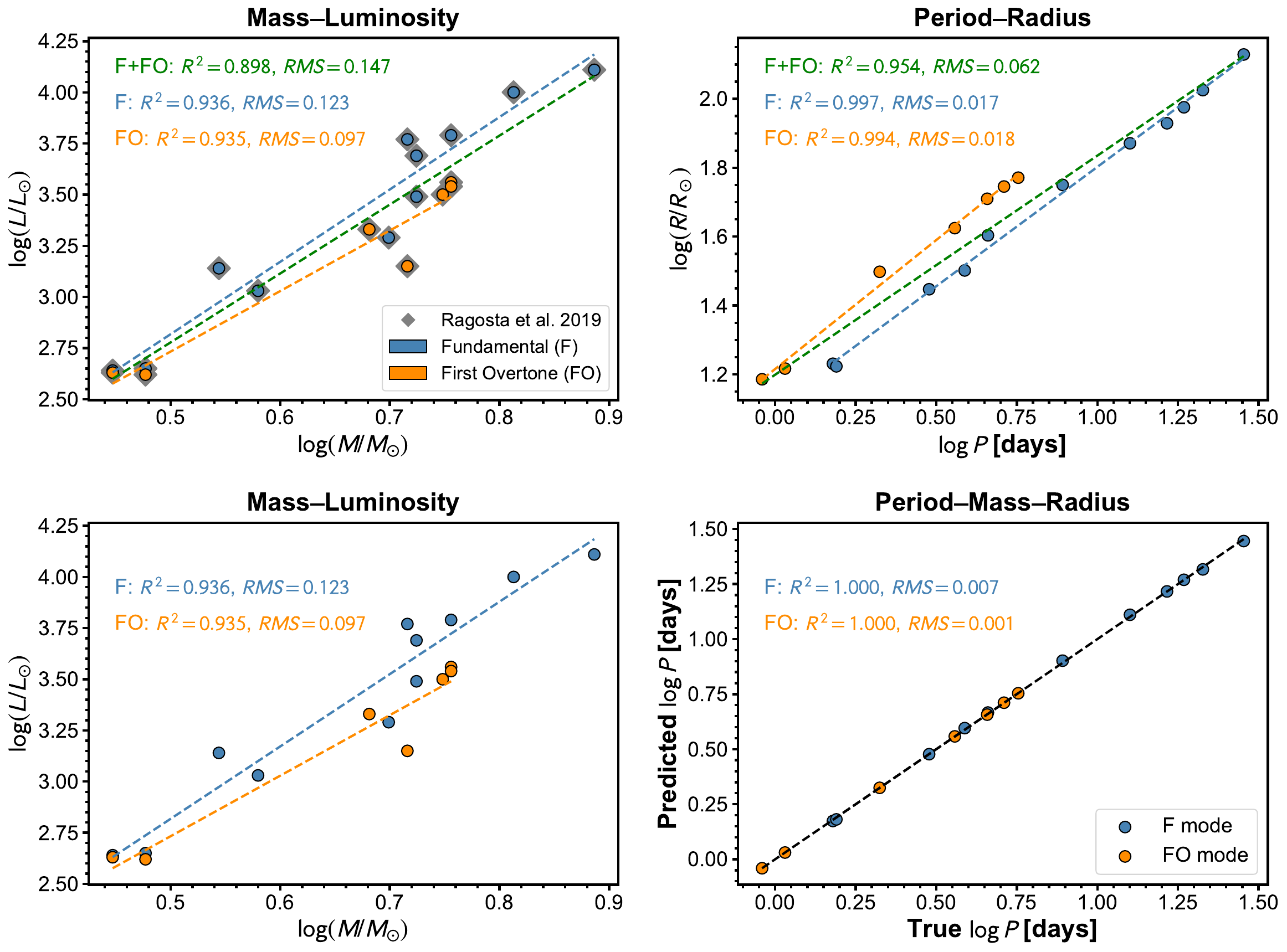}
\caption{This plot presents the ML, PR and PMR relations derived from our best-fitted model set. Two distinct relations for FU and FO cepheids are clearly distinct in each case.}
\label{fig:mlt_relations}
\end{figure*}

\begin{figure*}
\includegraphics[width=0.9\textwidth,keepaspectratio]{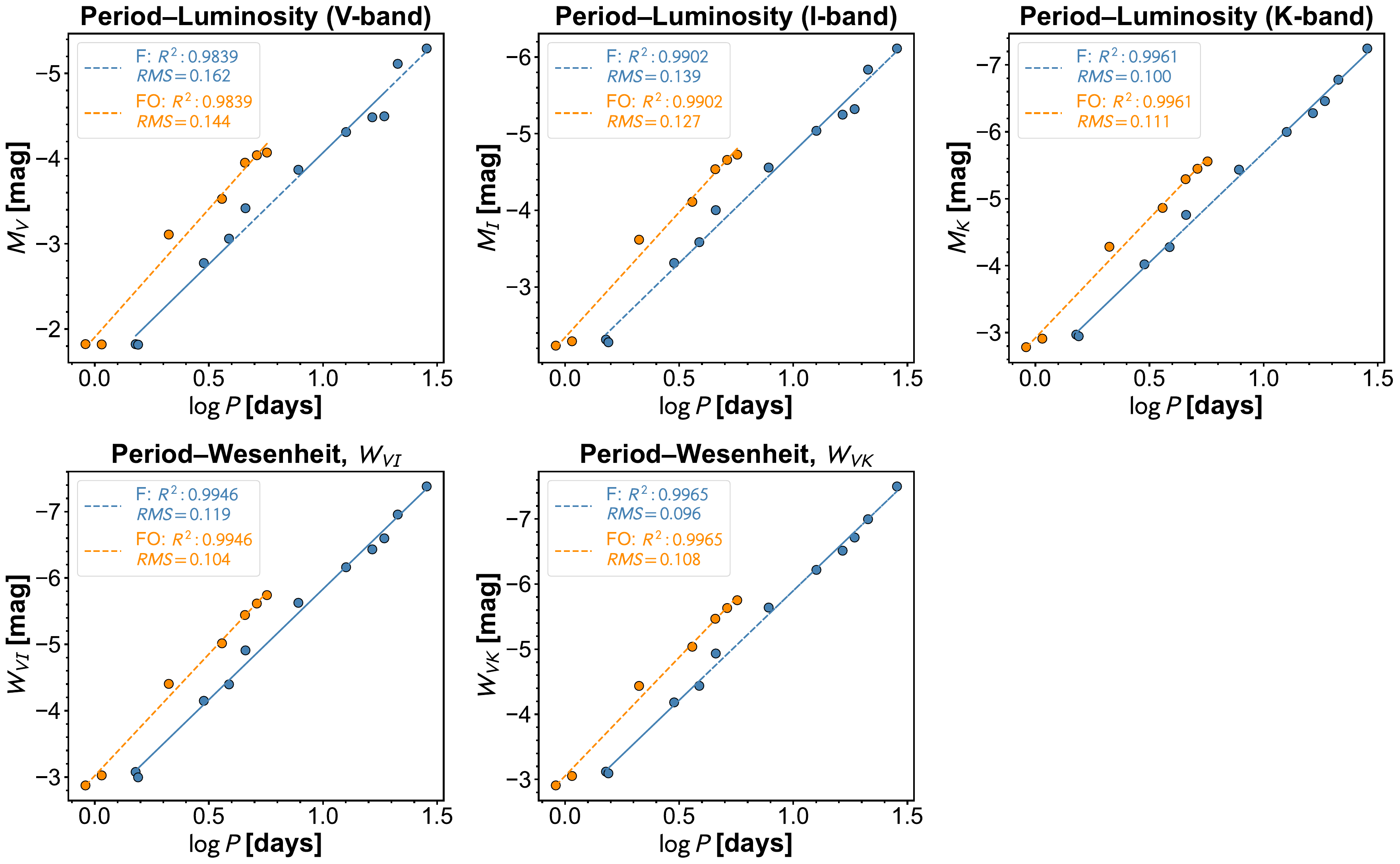}
\caption{This plot presents the PL and PW relations in $V,I$ and $K_{s}$-bands derived from out best-fitted model set. The relations are in very good agreement with that in R19.
}
\label{fig:pl_relations}
\end{figure*}

To quantify the pulsational properties of our model grid, we examined the relationships between key stellar parameters for both the FU and FO modes. In particular, we derived the period–radius (PR), period–mass-radius (PMR), and mass–luminosity (ML) relations shown in Fig.~\ref{fig:mlt_relations} as well as the PL relation in different bands shown in Fig.~\ref{fig:pl_relations}. Table~\ref{tab:relation_comparison}
provides a detailed comparison between the theoretical relations derived in the current study and those previously established by R19.

We did not observe any modifications in the masses and luminosities of the stars from those reported in R19. Additionally, the upper left panel of Fig.\ref{fig:mlt_relations} suggests that FO Cepheids exhibit more canonical behavior (i.e., less convective overshooting) than FU Cepheids, particularly at higher masses. 
R19 established a single PR relation for both FU and FO Cepheids by fundamentalizing FO Cepheids. However, in this work, we have derived two distinct relations for the two modes as shown in the upper right panel of Fig.\ref{fig:mlt_relations}. As expected, the PR relation from R19 aligns more closely with the FU relation derived in this study. Similarly, the PMR relation from R19 shows very good agreement with our FU-mode PMR relation.

Fig.~\ref{fig:pl_relations} presents the PL relations from our models in the $V,I$ and $K_{s}$-bands as well as the Wesenheit indices. These relations also show excellent agreement with those from R19, with comparable scatter.

\begin{table*}
\centering
\caption{Comparison of relations derived in this work with R19.}
\resizebox{\textwidth}{!}{%
\begin{tabular}{lcccccccc}
\hline
Types of Relation & Mode & \multicolumn{3}{c}{This Work} & \multicolumn{3}{c}{R19} \\
\cline{3-8}
 & & Slope & Intercept & RMS & Slope & Intercept & RMS \\
\hline
Mass-Luminosity &FU& $3.538 \pm 0.308$ & $1.049 \pm 0.037$ & 0.123 & -- & -- & -- \\
 & FO & $2.960 \pm 0.350$ & $1.253 \pm 0.037$ & 0.097 & -- & -- & -- \\
 & FU+FO & $3.369 \pm 0.283$ & $1.093 \pm 0.035$ & 0.147 & -- & -- & -- \\
\hline
Period-Radius &FU& $0.693 \pm 0.013$ & $1.110 \pm 0.005$ & 0.017 & -- & -- & -- \\
 & FO & $0.749 \pm 0.026$ & $1.216 \pm 0.007$ & 0.018 & -- & -- & -- \\
 & FU+FO & $0.636 \pm 0.035$ & $1.199 \pm 0.014$ & 0.062 & $0.70\pm0.02$ & $1.12\pm 0.01$ & $0.03$ \\
\hline
PL ($V$-band) &FU& $-2.618 \pm 0.112$ & $-1.451 \pm 0.107$ & 0.162 & $-2.63\pm0.11$ & $-1.54\pm0.11$ & 0.16 \\
 & FO & $-3.012 \pm 0.179$ & $-1.903 \pm 0.094$ & 0.144 & $-3.10\pm0.16$ & $-1.95\pm0.09$ & 0.13 \\
\hline
PL ($I$-band) &FU& $-2.898 \pm 0.096$ & $-1.863 \pm 0.092$ & 0.139 & $-2.93\pm0.11$ & $-1.95\pm0.10$ & 0.15 \\
 & FO & $-3.268 \pm 0.158$ & $-2.341 \pm 0.083$ & 0.127 & $-3.38\pm0.15$ & $-2.38\pm0.08$ & 0.12 \\
\hline
PL ($K_{s}$-band) &FU& $-3.277 \pm 0.068$ & $-2.410 \pm 0.066$ & 0.099 & $-3.30\pm0.09$ & $-2.43\pm0.09$ & 0.13 \\
 & FO & $-3.572 \pm 0.138$ & $-2.921 \pm 0.072$ & 0.111 & $-3.70\pm 0.14$ & $-2.90\pm0.07$ & 0.11 \\
\hline
PW ($W_{VI}$) &FU& $-3.333 \pm 0.082$ & $-2.500 \pm 0.079$ & 0.119 & $-3.39\pm0.10$ & $-2.58\pm0.09$ & 0.14 \\
 & FO & $-3.665 \pm 0.130$ & $-3.020 \pm 0.068$ & 0.104 & $-3.83\pm0.13$ & $-3.06\pm0.07$ & 0.11 \\
\hline
PW ($W_{VK_s}$) &FU& $-3.363 \pm 0.066$ & $-2.534 \pm 0.063$ & 0.096 & $-3.39\pm0.09$ & $-2.55\pm0.09$ & 0.13 \\
 & FO & $-3.644 \pm 0.134$ & $-3.053 \pm 0.070$ & 0.108 & $-3.78\pm0.13$ & $-3.02\pm0.07$ & 0.11 \\
\hline
Period-Mass-Radius &FU& \multicolumn{3}{c}{$\log P = -0.680\log \frac{M}{M_{\odot}} + 1.726\log \frac{R}{R_{\odot}} - 1.626$} & \multicolumn{3}{c}{--} \\
 & & \multicolumn{3}{c}{RMS = 0.007} & \multicolumn{3}{c}{} \\
 & FO & \multicolumn{3}{c}{$\log P = -0.648\log \frac{M}{M_{\odot}} + 1.666\log \frac{R}{R_{\odot}} - 1.707$} & \multicolumn{3}{c}{--} \\
 & & \multicolumn{3}{c}{RMS = 0.001} & \multicolumn{3}{c}{} \\
 & FU+FO & \multicolumn{3}{c}{$\log P = -1.612\log \frac{M}{M_{\odot}} + 2.180\log \frac{R}{R_{\odot}} - 1.809$} & \multicolumn{3}{c}{$\log P = -0.68\log \frac{M}{M_{\odot}} + 1.72\log \frac{R}{R_{\odot}} - 1.618$} \\
 & & \multicolumn{3}{c}{RMS = 0.058} & \multicolumn{3}{c}{RMS=0.005} \\
\hline
\end{tabular}
\label{tab:relation_comparison}
}
\tablefoot{``FU'' and ``FO'' denote the fundamental and first-overtone pulsation modes of Cepheids, respectively.}
\end{table*}

\section{Summary and conclusion}
\label{sec:conclusion}
We selected a sample of 18 Cepheids, including 11 FU and 7 FO Cepheids in the LMC, which were previously modeled by R19 using the Stellingwerf code. We obtained the light curve in the optical bands $(V, I)$ from the OGLE IV database and the NIR light curve $K_{s}-\text{band}$ from the VMC survey. We adopted the stellar
parameters -- mass, luminosity, effective temperature, and chemical compositions as reported in R19 and calibrated the TC parameter, namely $\alpha_\text{MLT},\alpha_m,\alpha_p$ and $\gamma_r$ to achieve the best agreement with the observed light curves. Our models successfully reproduced the observed light curves for all 18 stars in the sample, showing good agreement in magnitude shifts, radii, and phase alignment across multiple bands. The key findings of this study can be summarized as follows:

\begin{itemize}
\item We achieved good agreement with observations for all 18 stars. The derived stellar parameters — mass and luminosity are consistent with those reported by R19. However, for four stars, our models yield slightly different effective temperatures, with deviations ranging from approximately 20 to 250 K.
\item The morphology of the light curves is reproduced with the fiducial TC parameters of \mesarsp\ themselves. However, to achieve observation-consistent amplitude, we adjusted the mixing-length parameter ($\alpha_{\text{RSP}}^{\text{MLT}}$) and eddy viscous dissipation parameter ($\alpha_m$). We find that most of the stars with $T_{\text{eff}} > 6000$~K prefer the simplest convection theory, while stars with $T_{\text{eff}} < 6000$~K require the inclusion of turbulent pressure ($\alpha_p$), turbulent flux ($\alpha_t$), and in some cases, radiative loss ($\gamma_r$). We conclude that the fiducial TC parameters of \mesarsp\ are adequate for reproducing the general shape of light curves across the temperature range studied; however, parameter tuning is essential to accurately reproduce their observed amplitudes.
\item We also tested the inclusion of the parameters $\alpha_p$, $\alpha_t$, and $\gamma_r$ in models with $T_{\text{eff}} > 6000$~K. We found that most stars favor a low value of $\alpha_p$ (typically $\sim0.2$–$0.5$) and no radiative cooling (i.e., $\gamma_r = 0$). This could still lead to a degeneracy among the TC parameters when reproducing the observations. However, despite an excellent fit to the multi-band light curves, noticeable differences in the radial velocity amplitude remain, which could provide a means to break this degeneracy.
\item We also found good agreement in the multi-band PL, PR, PMR, and ML relations with R19. However, we have identified two distinct ML  relations for FU and FO Cepheids that had not been previously reported. This finding suggests a possible dependence on macroscopic phenomena that influence the ML relation.
\item We did not find any statistically significant correlation between $\alpha_\text{MLT}$ or $\alpha_m$ and the stellar parameters. However, we did observe subtle trends between $\alpha_\text{MLT}$ and $\alpha_m$ with the pulsation period and $T_\text{eff}$. We emphasize that before drawing any firm conclusions in this regard, a larger sample of observationally consistent models spanning the entire IS is required.
\item From this study, we also found that no single set of TC parameters can successfully model Cepheids across the entire IS. Even when restricting our analysis to observations from the central region of the IS, we observe that different combinations of TC parameters are required to reproduce the light curve features. This suggests that, rather than relying on a universal parameter set, it may be more effective to establish empirical or semi-empirical relations between the TC parameters and observable quantities. Such relations would significantly accelerate model computations and improve the efficiency of parameter space exploration. 
However, the number of observation-consistent models available at present is still statistically too small to robustly establish such relations.
\end{itemize}

In this work, we have not attempted to present a more physically motivated picture of Cepheids, nor do we claim that our models are more accurate in terms of stellar physics than those in previous studies. Rather, our primary goal was to reproduce the observed light curves and match the Stellingwerf models with \mesarsp\ models, focusing on light curve morphology and amplitudes.  Although \mesarsp\ is increasingly being adopted by the community \citep{das21,deka22b,das24,das25,kurb23,hocd24,espi25}, there remains a lack of sufficient non-linear models that successfully match observations.  However, developing such observation-consistent models is essential for constraining the effects of TC parameters on light curve morphology and the physical characteristics of Cepheids. These models also enable more precise determinations of stellar parameters, which are critical for galactic archaeology. Additionally, they provide a foundation for calibrating 1D TC prescriptions using results from 3D hydrodynamical simulations.

Thus far, our analysis has primarily focused on tuning the parameters $\alpha$ and $\alpha_m$ to achieve the desired amplitudes. \mesarsp\ performs quite well in reproducing the observed light curves of Cepheids without large modification in the $\alpha$ parameters. However, a comprehensive investigation of the other $\alpha$ parameters is still required to fully understand their influence on the minute details of the light curves. Additionally, the effects of these parameters on the dynamics of inner zones and the transition region remain largely unexplored -- not only in this work but also in the existing literature. We plan to address these aspects in our future studies.

\begin{acknowledgements}
We thank the anonymous referee for the constructive and valuable feedback.
MD acknowledges funding from the INAF 2023 Large Grant MOVIE (PI: Marcella Marconi) and project PRIN MUR 2022 (code 2022ARWP9C) ``Early Formation and Evolution of Bulge and HalO (EFEBHO)", PI: Marcella Marconi,  funded by European Union – Next Generation EU. MD also acknowledges the use of the High-Performance Computing facility Pegasus at IUCAA, India. To compute the models, we have used \textsc{MESA}~r23.05.1  \citep{paxt10,paxt13,paxt15,paxt18,paxt19}. G.D.S. acknowledges funding from the INAF-ASTROFIT fellowship (PI: G. De Somma), from Gaia DPAC through INAF and ASI (PI: M.G. Lattanzi), and from INFN (Naples Section) through the QGSKY and Moonlight2 initiatives. AB thanks the funding from the Anusandhan National Research Foundation (ANRF) under the Prime Minister Early Career Research Grant scheme (ANRF/ECRG/2024/000675/PMS). This research was supported by the International Space Science Institute (ISSI) in Bern/Beijing through ISSI/ISSI-BJ International Team project ID \#24-603 - “EXPANDING Universe” (EXploiting Precision AstroNomical Distance INdicators in the Gaia Universe). 
\end{acknowledgements}

\bibliographystyle{aa}
\bibliography{reference}

\begin{appendix}
\onecolumn

\section{Fiducial convection parameter sets in \mesarsp} \label{app:fiducialparameter}
We summarize the four combinations of the fiducial TC parameter sets as in \mesarsp\ below for the reader's reference.
\begin{table}[!h]
	\centering
	\caption{The fiducial turbulent convective parameter sets as reported in \citet{paxt19}.}
	\label{table:convective_set}
	\begin{tabular}{lcccccr} 
		\hline
        Parameters & set A &set B& set C & set D \\ \hline
        Mixing-length, $\alpha$ & $1.5$ & $1.5$ & $1.5$ & $1.5$ \\
        Eddy-viscous dissipation, $\alpha_{m}$ & $0.25$ & $0.50$ & $0.40$ & $0.70$ \\
        Turbulent source, $\alpha_{s}$ & $\frac{1}{2}\sqrt{\frac{2}{3}}$ & $\frac{1}{2}\sqrt{\frac{2}{3}}$ & $\frac{1}{2}\sqrt{\frac{2}{3}}$ & $\frac{1}{2}\sqrt{\frac{2}{3}}$ \\
        Convective flux, $\alpha_{c}$ & $\frac{1}{2}\sqrt{\frac{2}{3}}$ & $\frac{1}{2}\sqrt{\frac{2}{3}}$ & $\frac{1}{2}\sqrt{\frac{2}{3}}$ & $\frac{1}{2}\sqrt{\frac{2}{3}}$ \\
        Turbulent dissipation, $\alpha_{d}$  & $\frac{8}{3}\sqrt{\frac{2}{3}}$ & $\frac{8}{3}\sqrt{\frac{2}{3}}$& $\frac{8}{3}\sqrt{\frac{2}{3}}$ & $\frac{8}{3}\sqrt{\frac{2}{3}}$ \\
        Turbulent pressure, $\alpha_{p}$  & $0$ &  $0$ &  $\frac{2}{3}$ & $\frac{2}{3}$ \\
        Turbulent flux, $\alpha_{t}$  & $0$ &  $0$ & $0.01$ & $0.01$  \\
        Radiative cooling, $\gamma_{r}$  & $0$ &  $2\sqrt{3}$ & $0$ & $2\sqrt{3}$  \\
      \hline
	\end{tabular}
    \tablefoot{Set A is the simplest, while set B incorporates radiative cooling. set C further includes turbulent pressure and turbulent flux, and set D encompasses all these effects.}
\end{table}

\section{Best-fitted models}
\begin{figure*}
\includegraphics[width=0.9\textwidth,keepaspectratio]{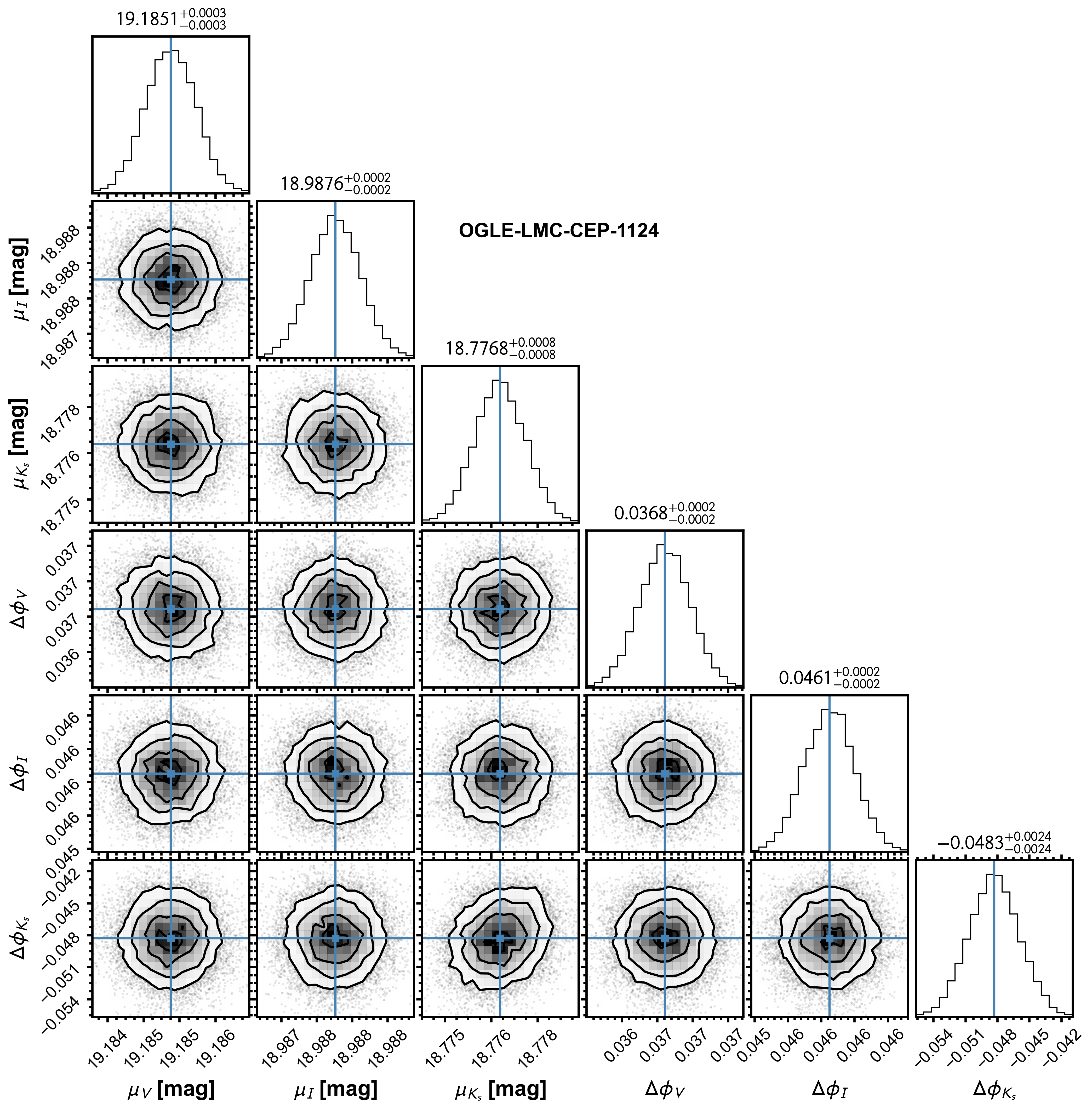}
\caption{An example of the posterior distribution from the Bayesian refinement analysis for the OGLE-LMC-CEP-1124 star showing the parameter co-
variances and the statistical uncertainties on the obtained parameters. }
\label{fig:corner}
\end{figure*}

We present an example corner plot of the posterior distribution obtained from the Bayesian refinement analysis in Fig.~\ref{fig:corner}.

\section{Best-fitting model parameters with set D equivalent TC parameters}
The following table presents the best-fitting model parameters for the plot shown in Fig.~\ref{fig:all_model}.

\begin{table*}[htbp]
\centering
\caption{Best-fitting model parameters, including mass, luminosity, effective temperature, and the \mesarsp\ TC parameters. }
\label{tab:bestfit_params}
\resizebox{\textwidth}{!}{%
\begin{tabular}{lccccccccccccccc}
\toprule
ID & P (days) & Mode & M ($M_{\odot}$) & $\log(L/L_{\odot})$ (dex) & $T_{\text{eff}}$ (K) & 
$\alpha^{\text{SW}}_{\text{ML}}$ & $\alpha^{\text{RSP}}_{\text{ML}}$ 
& $\alpha_m$ & 
$\alpha_{s}$ & $\alpha_c$ & $\alpha_d$ & $\alpha_p$ & 
$\alpha_{t}$ & $\gamma_r$ & $v_{\text{kick}}$\\
\midrule
1481 & 0.922 & FO & 3.00 & 2.62 & 6650 & 1.50 & 1.50 & 0.261 & 1.0 & 1.0 & 1.0 & 0.0 & 0.00 & 0.0 & 0.1\\
3131 & 1.095 & FO & 2.80 & 2.63 & 6450 & 1.80 & \textbf{1.60} & 0.250 & 1.0 & 1.0 & 1.0 & 0.0 & 0.00 & 0.0 & 0.1\\
3004 & 1.524 &FU& 3.00 & 2.65 & 6425 & 1.60 & 1.60 & 0.250 & 1.0 & 1.0 & 1.0 & 0.0 & 0.00 & 0.0 & 2.5\\
1523 & 1.572 &FU& 2.80 & 2.64 & 6425 & 1.60 & 1.60 & 0.240 & 1.0 & 1.0 & 1.0 & 0.0 & 0.00 & 0.0 & 2.0\\
3113 & 2.068 &FU& 5.20 & 3.15 & 6300 & 1.60 & \textbf{1.50} & 0.261 & 1.0 & 1.0 & 1.0 & 0.0 & 0.00 & 0.0 & 0.1\\
2138 & 3.011 &FU& 3.80 & 3.03 & 6175 & 1.70 & 1.70 & 0.260 & 1.0 & 1.0 & 1.0 & 0.0 & 0.00 & 0.0 & 0.1\\
3105 & 3.514 & FO & 4.80 & 3.33 & 6050 & 1.60 & 1.60 & 0.220 & 1.0 & 1.0 & 1.0 & 0.0 & 0.00 & 0.0 & 0.1\\
0961 & 3.711 &FU& 3.50 & 3.14 & \textbf{6250}(6100) & 1.50 & \textbf{1.70} & 0.230 & 1.0 & 1.0 & 1.0 & 0.0 & 0.00 & 0.0 & 1.5\\
1475 & 4.387 & FO & 5.60 & 3.50 & 6040 & 1.70 & \textbf{1.50} & 0.210 & 1.0 & 1.0 & 1.0 & 0.0 & 0.00 & 0.0 & 0.1\\
1124 & 4.457 &FU& 5.00 & 3.29 & \textbf{6060}(6040) & 1.70 & 1.70 & 0.400 & 1.0 & 1.0 & 1.0 & 0.0 & 0.00 & 0.0 & 0.1\\
1310 & 5.126 & FO & 5.70 & 3.54 & 5950 & 1.49 & \textbf{1.50} & 0.800 & 1.0 & 1.0 & 1.0 & 1.0 & 0.01 & 1.0 & 0.1 \\
0813 & 5.914 & FO & 5.70 & 3.56 & 5850 & 1.53 & \textbf{1.60} & 0.700 & 1.0 & 1.0 & 1.0 & 1.0 & 0.01 & 1.0 & 0.1\\
2012 & 9.934 &FU& 5.30 & 3.49 & 5775 & 1.60 & \textbf{1.80} & 0.410 & 1.0 & 1.0 & 1.0 & 1.0 & 0.01 & 0.0 & 0.1\\
1954 & 12.950 &FU& 5.30 & 3.69 & \textbf{5675}(5575) & 1.70 & 1.80 & 0.250 & 1.0 & 1.0 & 1.0 & 1.0 & 0.01 & 1.0 & 0.1\\
0546 & 15.215 &FU& 5.20 & 3.77 & 5575 & 1.70 & \textbf{1.90} & 0.250 & 1.0 & 1.0 & 1.0 & 1.0 & 0.01 & 1.0 & 0.1 \\
1086 & 15.360 &FU& 5.70 & 3.79 & 5450 & 1.50 & \textbf{2.30} & 0.250 & 1.0 & 1.0 & 1.0 & 1.0 & 0.01 & 1.0 & 0.1\\
2944 & 20.320 &FU& 6.50 & 4.00 & \textbf{5650}(5400) & 1.70 & \textbf{1.50} & 0.250 & 1.0 & 1.0 & 1.0 & 0.0 & 0.00 & 0.0 & 0.1\\
2019 & 28.103 &FU& 7.70 & 4.11 & 5425 & 1.70 & \textbf{2.00} & 0.300 & 1.0 & 1.0 & 1.0 & 1.0 & 0.01 & 1.0 & 0.1\\
\bottomrule
\end{tabular}%
}
\tablefoot{For comparison, we also include the mixing length parameter from the Stellingwerf code, $\alpha^{\text{SW}}_{\text{ML}}$, alongside the \mesarsp\ mixing length parameter, $\alpha^{\text{RSP}}_{\text{ML}}$. Parameters that differ slightly from those reported in R19 are highlighted in bold. The TC parameter values shown here are multiplicative factors applied to their corresponding base values shown in Table~\ref{table:convective_set}.}
\end{table*}

\section{Best-fitting model parameters with set D equivalent TC parameters}
\begin{figure*}[!htbp]
\includegraphics[width=0.95\textwidth,keepaspectratio]{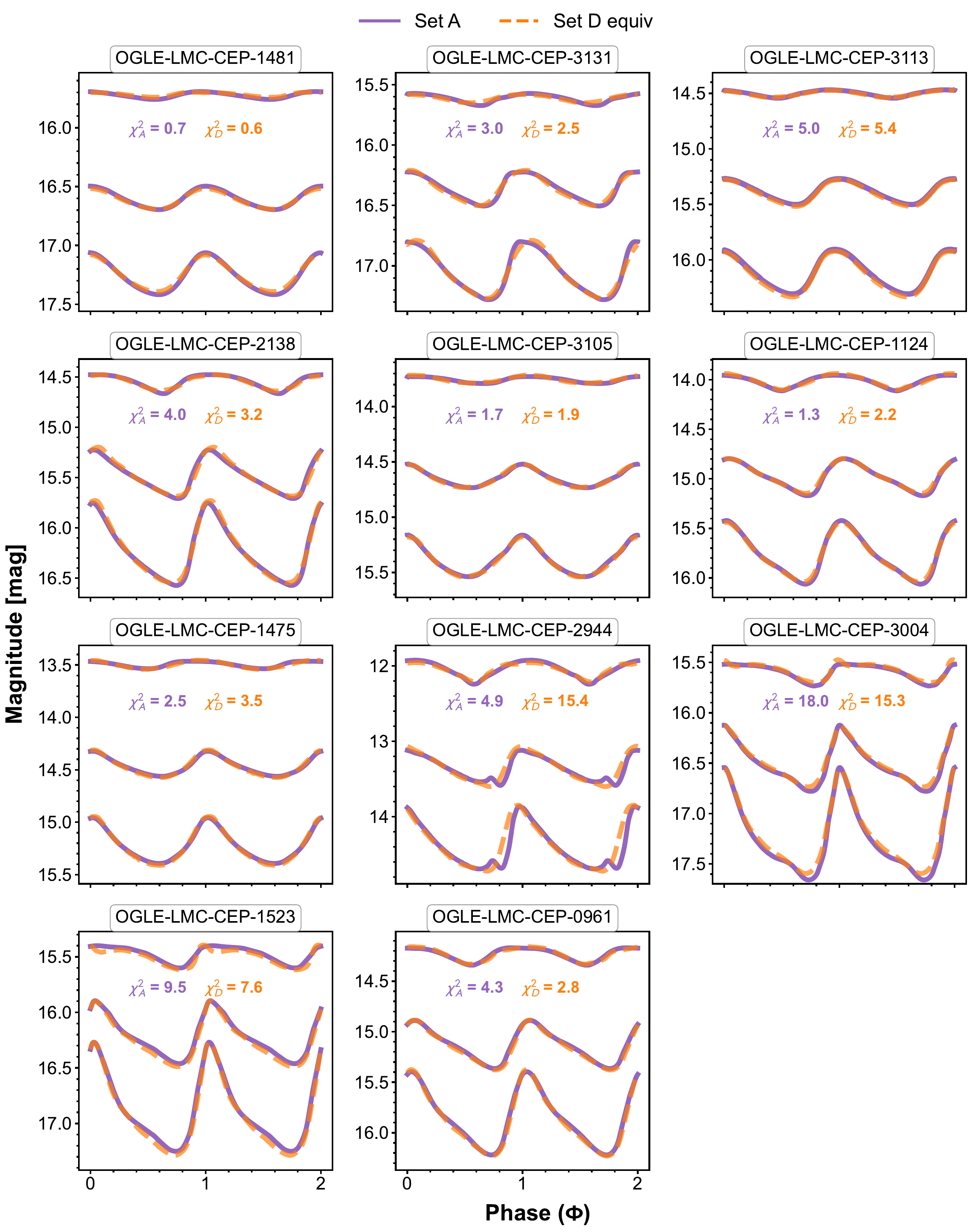}
\caption{The best-fitting models with the set D equivalent convection parameters are over-plotted with those with set A parameters.}
\label{fig:setA-setD_comp}
\end{figure*}

\begin{figure*}
\centering
\includegraphics[width=0.9\textwidth,keepaspectratio]{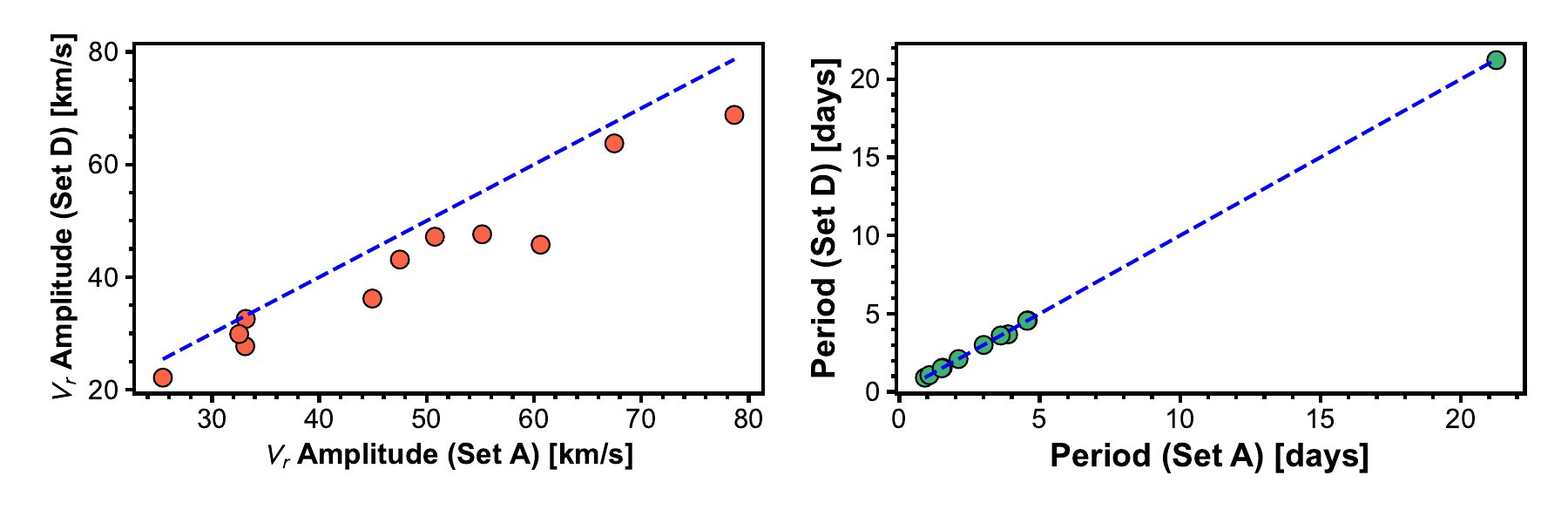}
\caption{Comparison of surface velocity amplitude (on the left) and period (on the right) between set A and set D equiv models. While the periods and light curve amplitudes show good agreement, some differences are noted in the surface velocity amplitudes.}
\label{fig:setA-setD_ample_comp}
\end{figure*}

The Table~\ref{tab:bestfit_params_setD} presents the best-fitting model parameters obtained after the inclusion of the TC parameters $\alpha_{p},\alpha_{t}$ and/or $\gamma_{r}$ in the convection theory. A comparison plot between set A and set D equivalent parameters and co-relation between periods and surface velocity amplitud are shown in Fig.~\ref{fig:setA-setD_comp} and Fig.~\ref{fig:setA-setD_ample_comp}, respectively. 

\begin{table*}
\centering
\caption{Best-fitting model parameters with equivalent set D TC parameters, including mass, luminosity, effective temperature, and the \mesarsp\ TC parameters. }
\label{tab:bestfit_params_setD}
\resizebox{\textwidth}{!}{%
\begin{tabular}{lccccccccccccccc}
\toprule
ID & P (days) & Mode & M ($M_{\odot}$) & $\log(L/L_{\odot})$ (dex) & $T_{\text{eff}}$\textbf{ }(K) & 
$\alpha^{\text{SW}}_{\text{ML}}$ & $\alpha^{\text{RSP}}_{\text{ML}}$ 
& $\alpha_m$ & 
$\alpha_{s}$ & $\alpha_c$ & $\alpha_d$ & $\alpha_p$ & 
$\alpha_{t}$ & $\gamma_r$ & $v_{\text{kick}}$\\
\midrule
1481 & 0.922 & FO & 3.00 & 2.62 & 6650 & 1.50 & 1.50 & 0.332 & 1.0 & 1.0 & 1.0 & 0.05 & 0.01 & 0.0 & 0.1\\
3131 & 1.095 & FO & 2.80 & 2.63 & 6450 & 1.80 & \textbf{1.60} & 0.250 & 1.0 & 1.0 & 1.0 & 0.1 & 0.01 & 0.0 & 0.1\\
3004 & 1.524 &FU& 3.00 & 2.65 & 6425 & 1.60 & 1.60 & 0.260 & 1.0 & 1.0 & 1.0 & 0.2 & 0.01 & 0.0 & 8.5\\
1523 & 1.572 &FU& 2.80 & 2.64 & 6425 & 1.60 & 1.60 & 0.240 & 1.0 & 1.0 & 1.0 & 0.0 & 0.00 & 0.0 & 6.5\\
3113 & 2.068 &FU& 5.20 & 3.15 & 6300 & 1.60 & \textbf{1.50} & 0.240 & 1.0 & 1.0 & 1.0 & 0.1 & 0.01 & 0.0 & 0.1\\
2138 & 3.011 &FU& 3.80 & 3.03 & 6175 & 1.70 & 1.70 & 0.510 & 1.0 & 1.0 & 1.0 & 0.6 & 0.01 & 1.0 & 20.0\\
3105 & 3.514 & FO & 4.80 & 3.33 & 6050 & 1.60 & 1.60 & 0.780 & 1.0 & 1.0 & 1.0 & 1.0 & 0.01 & 1.0 & 0.1\\
0961 & 3.711 &FU& 3.50 & 3.14 & \textbf{6250}(6100) & 1.50 & \textbf{1.70} & 0.190 & 1.0 & 1.0 & 1.0 & 0.2 & 0.01 & 0.0 & 1.5\\
1475 & 4.387 & FO & 5.60 & 3.50 & 6040 & 1.70 & \textbf{1.50} & 0.360 & 1.0 & 1.0 & 1.0 & 1.0 & 0.01 & 1.0 & 0.1\\
1124 & 4.457 &FU& 5.00 & 3.29 & \textbf{6060}(6040) & 1.70 & 1.70 & 0.430 & 1.0 & 1.0 & 1.0 & 0.1 & 0.01 & 0.0 & 0.1\\
1310 & 5.126 & FO & 5.70 & 3.54 & 5950 & 1.49 & \textbf{1.50} & 0.800 & 1.0 & 1.0 & 1.0 & 1.0 & 0.01 & 1.0 & 0.1 \\
0813 & 5.914 & FO & 5.70 & 3.56 & 5850 & 1.53 & \textbf{1.60} & 0.700 & 1.0 & 1.0 & 1.0 & 1.0 & 0.01 & 1.0 & 0.1\\
2012 & 9.934 &FU& 5.30 & 3.49 & 5775 & 1.60 & \textbf{1.80} & 0.410 & 1.0 & 1.0 & 1.0 & 1.0 & 0.01 & 0.0 & 0.1\\
1954 & 12.950 &FU& 5.30 & 3.69 & \textbf{5675}(5575) & 1.70 & 1.80 & 0.250 & 1.0 & 1.0 & 1.0 & 1.0 & 0.01 & 1.0 & 0.1\\
0546 & 15.215 &FU& 5.20 & 3.77 & 5575 & 1.70 & \textbf{1.90} & 0.250 & 1.0 & 1.0 & 1.0 & 1.0 & 0.01 & 0.0 & 0.1 \\
1086 & 15.360 &FU& 5.70 & 3.79 & 5450 & 1.50 & \textbf{2.30} & 0.250 & 1.0 & 1.0 & 1.0 & 1.0 & 0.01 & 1.0 & 0.1\\
2944 & 20.320 &FU& 6.50 & 4.00 & \textbf{5650}(5400) & 1.70 & \textbf{1.90} & 0.400 & 1.0 & 1.0 & 1.0 & 0.9 & 0.01 & 1.0 & 0.1\\
2019 & 28.103 &FU& 7.70 & 4.11 & 5425 & 1.70 & \textbf{2.00} & 0.300 & 1.0 & 1.0 & 1.0 & 1.0 & 0.01 & 1.0 & 0.1\\
\bottomrule
\end{tabular}%
}
\tablefoot{For comparison, we also include the mixing length parameter from the Stellingwerf code, $\alpha^{\text{SW}}_{\text{ML}}$, alongside the \mesarsp\ mixing length parameter, $\alpha^{\text{RSP}}_{\text{ML}}$. Parameters that differ slightly from those reported in R19 are highlighted in bold. The TC parameter values shown here are multiplicative factors applied to their corresponding base values shown in Table~\ref{table:convective_set}.}
\end{table*}

\end{appendix}

\end{document}